# Origin of Hole-Trapping States in Solution-Processed Copper(I) Thiocyanate (CuSCN) and Defect-Healing by $I_2$ Doping

Pimpisut Worakajit,[1] Pinit Kidkhunthod,[2] Saran Waiprasoet,[1] Hideki Nakajima,[2] Taweesak Sudyoadsuk,[1] Vinich Promarak,[1] and Pichaya Pattanasattayavong[1,*]

[1] Department of Materials Science and Engineering, School of Molecular Science and Engineering, Vidyasirimedhi Institute of Science and Technology (VISTEC), Rayong, 21210, Thailand

[2] Synchrotron Light Research Institute (Public Organization), 111 University Avenue, Muang, Nakhon Ratchasima, 30000, Thailand

*Correspondence: pichaya.p@vistec.ac.th

**Abstract**

Solution-processed copper(I) thiocyanate (CuSCN) typically exhibits low crystallinity with short-range order; the defects result in a high density of trap states that limit the device performance. Despite the extensive electronic applications of CuSCN, its defect properties have not been studied in detail. Through X-ray absorption spectroscopy, pristine CuSCN prepared from the standard diethyl sulfide-based recipe is found to contain under-coordinated Cu atoms, pointing to the presence of SCN vacancies. A defect passivation strategy is introduced by adding solid $I_2$ to the processing solution. At small concentrations, the iodine is found to exist as $I^-$ which can substitute for the missing $SCN^-$ ligand, effectively healing the defective sites and restoring the coordination around Cu. Applying $I_2$-doped CuSCN as a p-channel in thin-film transistors shows that the hole mobility increases by more than five times at the optimal doping concentration of 0.5 mol%. Importantly, the on/off current ratio and the subthreshold characteristics also improve as the $I_2$ doping method leads to the defect healing effect while avoiding the creation of detrimental impurity states. An analysis of the capacitance-voltage characteristics corroborates that the trap state density is reduced upon $I_2$ addition. The contact resistance and bias-stress stability of the devices also improve. This work shows a simple and effective route to improve hole transport properties of CuSCN which is applicable to wide-ranging electronic and optoelectronic applications.

1. Introduction

Copper(I) thiocyanate (CuSCN) has arisen as a novel 3D coordination polymer (CP) with promising characteristics such as attractive hole transport properties, high optical transparency, and solution processability.[1–3] As a result, a variety of CuSCN-based optoelectronic devices have been demonstrated, particularly organic solar cells (OSCs),[4–6] perovskite solar cells (PSCs),[7–9] organic light emitting diode (OLEDs),[10–12] and thin film transistors (TFTs).[13–15] The breadth and consistency of device demonstrations establishes CuSCN as a unique CP-based semiconductor for practical applications, and improvements in the properties of CuSCN can have a wide-ranging impact for its use.

CuSCN is typically solution-processed at relatively low temperatures, resulting in low crystallinity which likely contains a high degree of structural disorder.[13,16,17] Understanding the defect properties in order to engineer them proves a highly successful strategy in the developments of semiconductors and electronic devices.[18–21] Indeed, a high density of tail states and hole-trapping states is known to exist in CuSCN,[22–24] but their origin has not been elucidated. It has been shown that the p-type conductivity of CuSCN increases under Cu-deficient condition and decreases under Cu-rich condition.[25] Accordingly, theoretical studies show that the two basic native defects, Cu vacancies ($V_{Cu}$) and SCN vacancies ($V_{SCN}$), should result in acceptor and donor states, respectively.[1,26,27] As the acceptor states receive electrons from the valence band of CuSCN, they increase the hole concentration. In contrast, the creation of donor states compensates for the acceptors, acting as hole killers, hole traps, or hole scattering centers.[28,29] Therefore, the presence of compensating donor states would present a serious problem to the hole transport in p-type CuSCN, and their passivation is therefore of high importance.

To improve the hole transport properties of CuSCN, doping is one of the main approaches generally explored. The solution processability of CuSCN allows the doping procedure to be carried out simply by mixing dopants to the CuSCN solution. For example, molecular dopants with low-lying lowest unoccupied molecular orbital (LUMO) energy levels have been investigated. A fluorinated fullerene derivative $C_{60}F_{48}$ was shown to be effective at p-doping CuSCN, employed as a p-channel layer in TFTs (optimal doping level = 0.5 mol%) and a hole-transport layer (HTL) in OSCs (optimal doping level = 0.005 – 0.01 mol%), respectively.[30] Improved device performance was evident from higher field-effect hole mobility, bias-stress stability, solar cell fill factor (FF) and power conversion efficiency (PCE) that could be attributed to increased hole concentration and better interfacial uniformity. Similarly, 2,3,5,6-tetrafluoro-7,7,8,8-tetracyanoquinodimethane (F4-TCNQ) was used as a dopant for a CuSCN HTL in PSCs (optimal doping level = 0.03 wt%, equivalent to 0.013 mol%); the results showed a closer energy alignment between the CuSCN HTL and the active perovskite layer, yielding improvements across all solar cell metrics.[31]

A closely related compound, copper(I) iodide (CuI), is also emerging as a promising solution-processable p-type inorganic semiconductor.[32–34] CuI has been employed to dope CuSCN or form composites with CuSCN as well. CuI doping of electrolyte-gated CuSCN TFTs (doping level = 5 wt%, equivalent to 3.3 mol%) showed an increase in the field-effect hole mobility.[35] The improvement was attributed to a p-doping effect that increased the hole concentration and the co-existence of the CuI phase which is reported to have higher hole mobility.[36] CuSCN-CuI composites were investigated as a HTL for polymer-based OLEDs (optimal CuI concentration = 25 wt%, equivalent to 17.6 mol%). The enhancement in device performance was ascribed to higher hole mobility due to the coincident CuI phase and better energy matching between the HTL and the emissive layer.[37]

The abovementioned examples rely primarily on the p-doping effect, i.e., increasing the hole concentration and using the excess holes to overcome the trapping states (trap filling).[38] While the current and hole mobility may apparently increase, other device characteristics could suffer in a trade-off. Particularly for the TFT application, the p-doping effect leads to detrimental increases in the off current and subthreshold swing,[30,35] due to higher hole concentration and higher tail state density from the dopants. The increase in the off current is typically larger than the gain in the on current, hence reducing the on/off current ratio. Herein, we report a doping route based on solid $I_2$ which is processable in the standard diethyl sulfide (DES)-based solution for CuSCN TFT device fabrication. Crucially, we show that an optimal amount of $I_2$ leads to a defect-healing effect that remarkably improves the hole transport properties of CuSCN and device performance in all metrics. By conducting the synchrotron-based X-ray absorption spectroscopy, we reveal that the Cu centers in the pristine CuSCN sample are under-coordinated, suggesting the presence of $V_{SCN}$. At low $I_2$ doping amounts, the iodine is found to exist as $I^-$ that can fill in the voids of $V_{SCN}$ and complete the coordination around Cu. At the optimal doping concentration of 0.5 mol%, the field-effect hole mobility increases by more than five times. Importantly, the on/off current ratio also increases to more than $10^4$, which is distinct from previous doping attempts. The $V_{SCN}$ defect passivation by $I^-$ is also reflected in the decrease in the trap state density, evident from the reduction in the TFT subthreshold swing as well as from a detailed capacitance-voltage characterization. The optimally doped devices also exhibit a lower contact resistance and superior bias-stress stability.

## 2. Chemical and Microstructural Analyses of $I_2$-doped CuSCN

$I_2$ doping of CuSCN has been attempted previously; however, the reported method proceeded by treating CuSCN films with halogen gases,[39] which carried risks and setup complications. The method reported herein is very simple; solutions of $I_2$ beads and CuSCN powder, each dissolved separately using DES as a solvent at 10 mg ml$^{-1}$, were mixed with different volume ratios to yield various concentrations of $I_2$ doping (Table S1 and full experiment details in Supporting Information). Hereafter, the samples are designated by their doping concentrations or as 'pristine' for undoped CuSCN. The color of $I_2$-doped CuSCN solutions changed from colorless to brown when $I_2$ concentration was equal to or higher than 5 mol% as shown in Figure S1. The color of drop-cast samples also changed from greyish white to brown at the same concentration. However, spin-coated samples retained the fully transparent appearance of CuSCN thin films following the annealing step (data shown in Section 4).

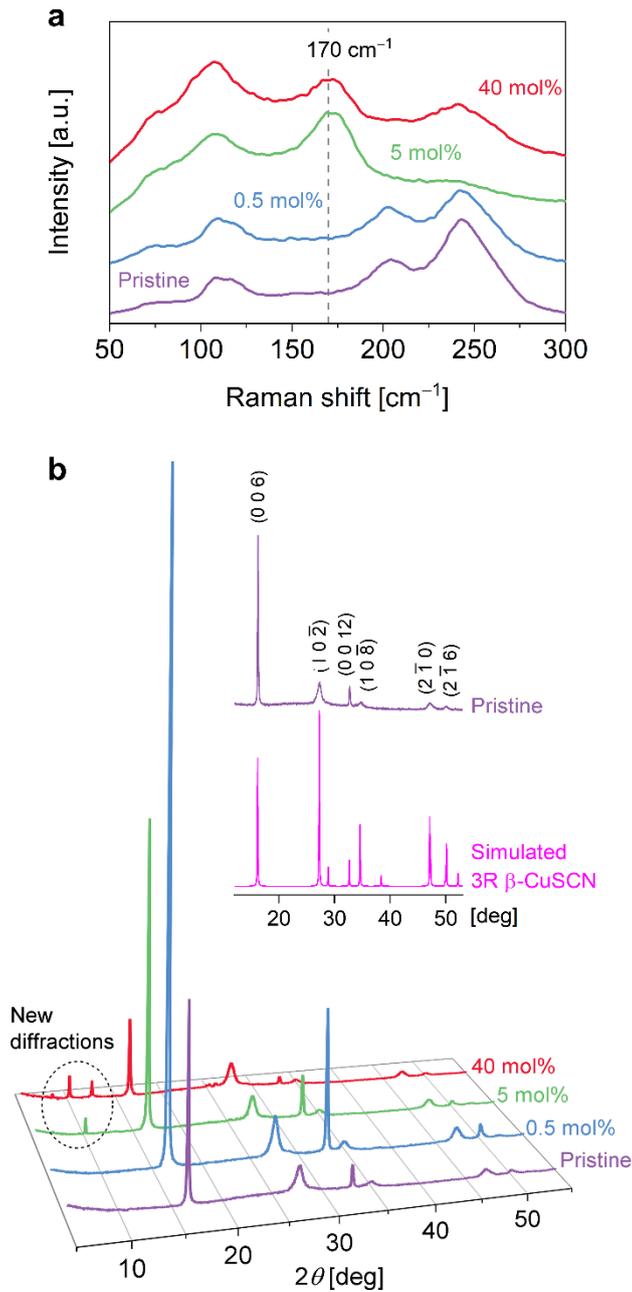

**Figure 1.** (a) Raman spectra. The peak at 170 cm$^{-1}$ is ascribed to I$_2$ acting as an electron acceptor. (b) X-ray diffraction data of the pristine and I$_2$-doped CuSCN samples. (b, inset) The diffractions of the pristine sample can be matched with those from the 3R β-CuSCN polytype.

Raman spectroscopy was used to investigate the influence of the I$_2$ dopant on the chemical structure and bonding properties of CuSCN. Samples prepared by drop-casting were employed in this case (see Supporting Information). The Raman spectra of pristine and doped samples at 0.5, 5, and 40 mol% recorded in the wavenumber range of 50 to 3500 cm$^{-1}$ are shown in Figure S2. The bands at 204, 243, 432, 747 and 2173 cm$^{-1}$ are characteristics of CuSCN and can be assigned to Cu-S stretching, Cu-N stretching, SCN bending, C-S stretching, and C≡N stretching, respectively.[40–42] With the increasing I$_2$ doping concentration, a significant change can be observed in the low wavenumber region. Specifically, **Figure 1**a shows a band at about 170 cm$^{-1}$ appearing for 5 and 40 mol% doping conditions. For solid I$_2$, a Raman signal is found at 180 cm$^{-1}$, and the shift to lower wavenumbers is attributed to molecular I$_2$ acting as an electron acceptor (partially reduced),[43,44] which in our case can be coupled with Cu$^+$ as a possible electron donor. The X-ray absorption data, discussed in the next section, corroborates this analysis. We note that the molecular I$_2$ character was not observed in the 0.5 mol% doping condition (optimal doping level for TFTs as shown later).

To investigate the crystalline phases, we carried out X-ray diffraction (XRD) of the drop-cast samples. First, the results of the pristine sample and reference CuSCN (powder, 99%, Sigma-Aldrich, also used for CuSCN solution preparation) were compared against data of the reported structures. As shown in Figure S3, the reference CuSCN powder was predominantly of the 3R β-CuSCN polytype (ICSD 24372)[45] with a small presence of α-CuSCN (ICSD 124).[46] The pristine drop-cast sample exhibited lower crystallinity, and only the 3R β-CuSCN phase can be identified with diffractions at $2\theta$ = 16.2°, 27.3°, 32.7°, 34.7°, 47.2°, and 50.0° associated with (0 0 6), (1 0 $\bar{2}$), (0 0 12), (1 0 $\bar{8}$), (2 $\bar{1}$ 0), and (2 $\bar{1}$ 6) planes (Figure 1b, inset). The preferred orientation along the crystalline c-axis of the pristine CuSCN sample is noted from the highest intensity detected from the (0 0 6) diffraction. Upon the addition of I$_2$, the XRD pattern changed as displayed in Figure 1b. The six peaks of 3R β-CuSCN were still present, suggesting that the dominant crystalline phase remained the same. At 0.5 mol% I$_2$ concentration, the c-axis orientation became even more pronounced, with very strong intensities from the (0 0 6) and (0 0 12) planes. With a further increase in $I_2$, the degree of the preferred orientation decreased but persisted. At 5 mol% and 40 mol% concentrations, new diffractions were observed to arise, evident from a small peak at $2\theta = 9.9°$ and also two additional peaks at $2\theta = 8.1°$ and $12.2°$ for the latter (40 mol%). The new diffractions at high doping concentrations cannot be conclusively identified at this point, but we speculate that they were resulted from the inclusion of molecular $I_2$ in the CuSCN structure as discussed in the next section.

## 3. Local Atomic Structures and Defect Healing by $I_2$ Doping

The effects of $I_2$ addition on the local atomic structures of CuSCN were investigated in detail by X-ray absorption spectroscopy (XAS). The samples were similar to those used for Raman and XRD experiments. The XAS measurements of the Cu K-edge and I L3-edge were conducted at the beamline BL5.2 SUT-NANOTEC-SLRI, Synchrotron Light Research Institute (SLRI), Thailand. The XAS spectra were recorded in both the X-ray absorption near edge structure (XANES) and the extended X-ray absorption fine structure (EXAFS) spectral ranges. The standard samples for reference were CuSCN powder (99%, Sigma-Aldrich), CuI powder (99.5%, Sigma-Aldrich), and $I_2$ beads (99.9%, Wako).

The XANES results of the Cu K-edge are shown in Figure S4a, and specifically those of pristine, 0.5 mol% $I_2$-doped (optimal condition for TFTs), and CuSCN standard samples are displayed in **Figure 2**a for clarity. For all samples, the rising edge showed a feature at 8984 eV associated with the 1s → 4p electronic transition of $Cu^+$; no pre-edge features at energies <8980 eV related to 1s → 3d transition of $Cu^{2+}$ were detected.[47,48] This indicates that Cu centers were in the $d^{10}$ $Cu^+$ state (no 3d holes) and $I_2$ doping did not alter the oxidation state of Cu. The overall spectra of all samples were generally consistent with a 4-coordinate $Cu^+$ (Ref.[47,48]) and similar to previously reported data of CuSCN.[49,50] The white line of the samples, which is a strong absorption peak after the edge, was located at 8997 eV. Interestingly, we observe that the pristine CuSCN sample exhibited a noticeably lower white line intensity than the CuSCN standard, indicating a slightly lower coordination number of the Cu atoms in the pristine sample.[51–53] The incomplete coordination around Cu$^+$ centers by the SCN$^−$ ligands signifies that the solution-processing of CuSCN results in the formation of SCN vacancies ($V_{SCN}$). Remarkably, after adding 0.5 mol% of $I_2$, the white line intensity and the overall spectrum matched with that of the CuSCN standard. A further increase in $I_2$, especially at 40 mol%, led to more deviations from the spectrum of the CuSCN standard. This suggests that at the optimal doping amount, $V_{SCN}$ defects are passivated, leading to a complete tetrahedral coordination around Cu atoms. However, an excessively high doping amount leads to some disruption to the coordination sphere around Cu atoms that can be attributed to the large-sized molecular $I_2$.

Figure S4b shows the Fourier-transformed EXAFS data in the R-space of the Cu K-edge X-ray absorption spectra. The CuSCN standard exhibited a major peak at 1.84 Å, which can be ascribed to the scattering path of the dominant Cu-S bonding.[54] The broadening of this peak may obscure the signal from the Cu-N scattering path, typically observed around 1.5 Å. For the pristine sample, the amplitude of the main peak was significantly reduced when compared to that of the CuSCN standard. This can be attributed to a diminishing degree of bonding in the first coordination sphere around Cu,[55] consistent with some deficiency of SCN$^−$ ligands as discussed above. As highlighted in Figure 2b, the addition of 0.5 mol% $I_2$ to the processing solution led to the recovery of EXAFS data to a similar profile to that of the CuSCN standard. This further corroborates that an optimal amount of $I_2$ can heal the defect sites around Cu and improve the crystallinity. However, an excessive amount disrupts the bonding around Cu, particularly evident from the 40 mol% doping concentration (Figure S4b).

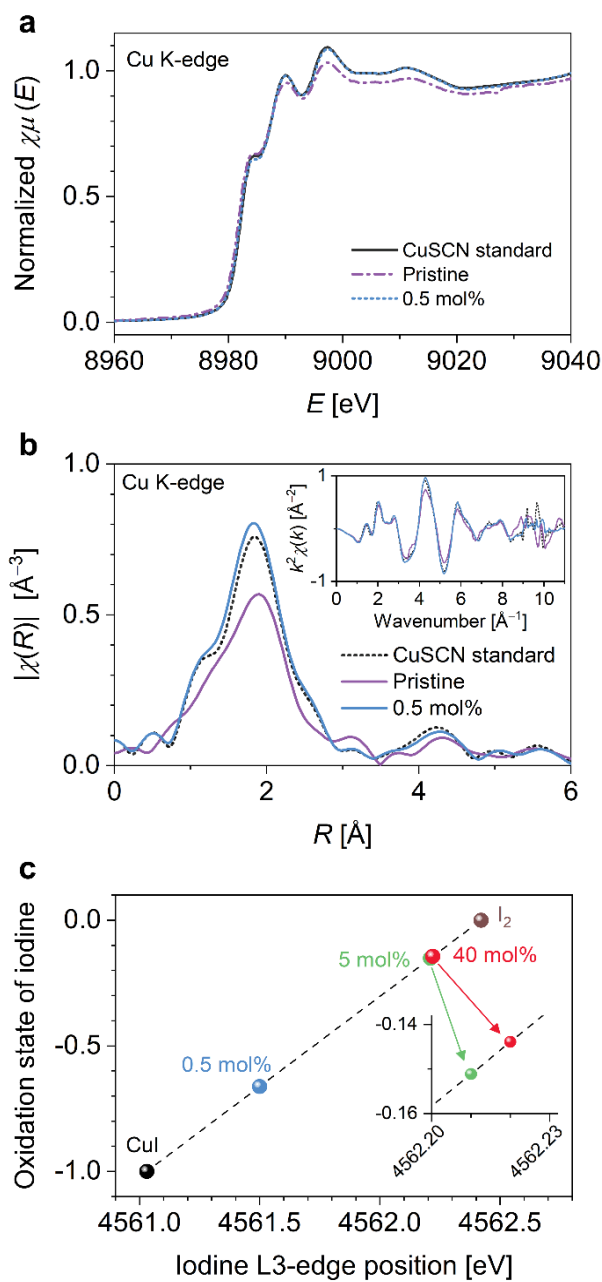

**Figure 2.** Cu K-edge (a) XANES spectra and (b) EXAFS spectra in R-space (inset, in k-space). The radial distance (R) is shown without phase shift correction. (c) Oxidation state of iodine determined from the absorption edge positions of I L3-edge XANES.

In addition to the Cu K-edge data, the analysis of I L3-edge XAS measurements allowed us to elucidate the oxidation state and local environment of the added iodine. The XANES spectra of CuSCN samples with various $I_2$ concentrations are shown in Figure S5a. The absorption edges of CuI and $I_2$ standards were located at 4561.0 and 4562.4 eV, which can be assigned to iodine in the oxidation state -1 and 0, respectively. The L3-edges of the $I_2$-doped samples were located between the two standards, and by using a linear correlation analysis the oxidation state of iodine can be found as shown in Figure 2c:[56] -0.67 for the 0.5 mol% sample and around -0.15 for the 5 and 40 mol% samples. This result confirms that, at 0.5 mol%, iodine exists as $I^-$ ligands which can replace the missing $SCN^-$ ligands, hence passivating $V_{SCN}$ as mentioned above. The slightly higher oxidation state than -1 is ascribed to unequal charge distribution between $I^-$ and the other $SCN^-$ ligands that surround the $Cu^+$ center. In contrast, when the doping amount was too high, iodine likely exists as $I_2$. The slightly reduced oxidation state agrees with the Raman spectroscopy results that showed molecular $I_2$ acting as an electron acceptor in the highly doped samples.

In addition, Figure S5b shows the I L3-edge EXAFS spectra in the R-space of $I_2$-doped samples. The CuI and $I_2$ standards showed a major peak around 2.2 Å and 2.6 Å, which can be ascribed to Cu-I and I-I scattering paths respectively.[57,58] Compared to the CuI standard, the sample with 0.5 mol% $I_2$ exhibited a similar peak for the first Cu-I scattering but with an absence of the second peak at 3.8 Å. The latter indicates that the outer coordination sphere of iodine in this sample was different from that in CuI; this further substantiates that $I^-$ was filling in for $V_{SCN}$ in the structure of CuSCN. This result is consistent with the XRD data which did not show the presence of the CuI phase. The EXAFS spectra at higher doping concentrations of 5 and 40 mol% showed different profiles to both CuI and $I_2$ standards. From Raman and XANES analyses, molecular $I_2$ likely existed in these samples; the EXAFS data then supplements that they did not exist as solid $I_2$ clusters but were included into the main CuSCN structure which led to a different local environment when compared to that of the $I_2$ standard.

By combining the Raman spectroscopy, XRD, and XAS findings, we deduce that adding a small amount of $I_2$ (e.g., 0.5 mol%) to the CuSCN processing solution leads to the *defect healing* effect through $I^-$ filling in the vacancies of $SCN^-$, passivating the structural defects and improving the crystallinity. $I^-$ is well-known to act as a ligand that coordinates with $Cu^+$ centers in coordination polymers or complexes.[59,60] The ionic radius of $I^-$ (220 pm)[61] is comparable to that of $SCN^-$ (215-220 pm).[62] Furthermore, the two anions can similarly stabilize Cu in the +1 oxidation state, emphasizing their chemical compatibility. However, when the doping concentration is too high (e.g., $\geq$ 5 mol%), iodine exists as molecular $I_2$ units. Although they still coordinate to $Cu^+$ centers, their large size leads to more disorder around $Cu^+$ and decrease the overall crystallinity.

## 4. Thin-Film Morphology, Optical Properties, and Electronic Energy Levels

The effects of $I_2$ addition on the surface morphology of CuSCN thin films at various doing concentrations were investigated by atomic force microscopy (AFM), and the topographic images are shown in Figure S6a-f. The pristine film exhibited the nanoscale elongated grain characteristic similar to previous reports.[17,63] Upon doping with $I_2$, the overall feature remained broadly similar but with some varying roughness and grain size. Figure S6g displays the surface height ($Z_{AFM}$) histograms of the samples. The root-mean-square roughness ($\sigma_{RMS}$) and the average grain diameter ($D_{avg}$) from the analysis are reported in Table S2 and plotted in Figure S6h. The trend in the morphology change was not obvious, except for 40 mol% doping level in which the grains became noticeably smaller. This is possibly due to the disruption of the crystallinity by the large molecular $I_2$ species in the structure as discussed in the previous section.

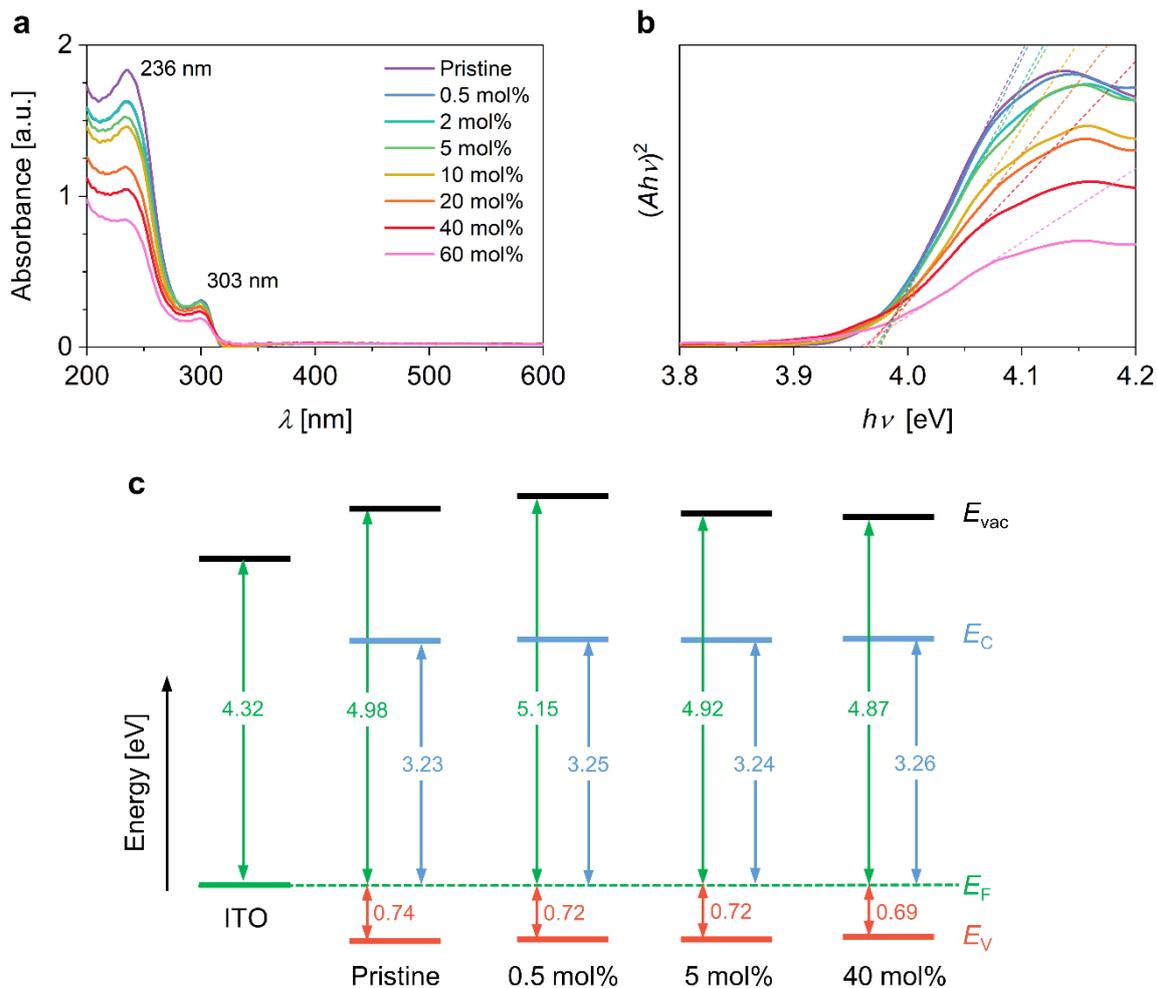

**Figure 3.** (a) Optical absorbance spectra. (b) Tauc plots for the determination of the optical band gaps. (c) Energy diagrams constructed with data from photoelectron spectroscopy (PES) and the optical band gaps. $E_v$: valence band edge; $E_F$: Fermi level; $E_c$: conduction band edge; $E_{vac}$: vacuum level.

The optical properties of the films were studied by the ultraviolet-visible-near infrared (UV-Vis-NIR) spectroscopy. Following the thin-film fabrication by spin-coating and annealing at 100 °C, all films became transparent. The measured optical transmittance and reflectance of pristine and $I_2$-doped films (on fused silica substrates) are shown in Figure S7a-b. In the 400-1400 nm wavelength region, all films demonstrated an average transparency higher than 95%.

The converted absorption spectra are displayed in **Figure 3**a. All samples show similar absorption peaks at 236 and 303 nm, which are characteristics of CuSCN.[30,63] The increase in the transmittance and the corresponding decrease in the absorbance with the increasing doping concentration are attributed to the reduction in the film thickness (Figure S8). The optical band gaps ($E_g^{opt}$) of the samples were determined from the Tauc plots as presented in Figure 3b and found to be in the range of 3.95-3.97 eV (Table S3). This result suggests that $I_2$ doping does not affect the optical properties of CuSCN films in general.

To study the electronic energy levels of the samples, photoelectron spectroscopy (PES) was performed using synchrotron radiation at the Beamline BL3.2Ua, Synchrotron Light Research Institute (SLRI), Thailand.[64] Figure S9 shows the PES spectra of pristine and $I_2$-doped CuSCN films (0.5, 5, and 40 mol%) prepared on ITO substrates, and the resulting energy levels analyzed from the data are reported in Table S4. The spectra exhibit similar characteristics to the previous report of CuSCN.[17,30] The PES data combined with the $E_g^{opt}$ from the optical measurements allow the construction of the energy diagrams as illustrated in Figure 3c. We can observe that with respect to the reference Fermi level ($E_F$) the band edge positions did not vary significantly, nor did they exhibit clear trends. No additional states from doping were observed because our process of adding $I_2$ does not lead to a strong p-doping effect but rather a defect-healing effect. Some variations in the work functions (i.e., $E_F$ with respect to the vacuum level or $E_{vac}$) are likely resulted from the changes in the crystallinity and orientation of the microstructure.[65–67]

## 5. Thin-film Transistors and Analysis of Hole Transport Properties

Thin-film transistors (TFTs) based on the pristine and $I_2$-doped CuSCN layers were fabricated to elucidate the effects of doping on the hole transport properties. Details of the fabrication and characterizations can be found in the Supporting Information. The representative transfer and output characteristics of all samples are included in Figure S10 and S11. For clarity of comparison, only the forward-sweep transfer curves of pristine, 0.5, 5 and 40 mol% I$_2$-doped conditions operated in the saturation regime are shown in **Figure 4**a. The results reveal that the p-type characteristics of CuSCN transistors were greatly improved by I$_2$ doping at 0.5 mol% condition. The on drain current ($I_{D,on}$) was increased by one order of magnitude with a small increase in the off drain current ($I_{D,off}$). The same data in Figure 4a is plotted as $\sqrt{I_D}$ versus the gate voltage ($V_G$) in Figure 4b, clearly showing the positive shift of the threshold voltage ($V_{th}$) and the increase in the saturation hole mobility ($\mu_{sat}$) which are reflected in the x-intercept and slope, respectively. Full TFT device parameters are reported in Table S5. For each condition, at least 10 devices were measured, and the average and uncertainty (from the arithmetic mean and standard deviation) were computed.

The evolution of $\mu_{sat}$, $V_{th}$, and the subthreshold swing ($S_{th}$) with I$_2$ concentration are visualized in Figure 4c. We can clearly see that at the optimal concentration of 0.5 mol%, the device characteristics were drastically improved from the undoped condition: the average $\mu_{sat}$ increasing from $1.3\times10^{-2}$ to $7.1\times10^{-2}$ cm$^2$ V$^{-1}$ s$^{-1}$; the average $V_{th}$ shifting from -4.6 to -0.4 V; and the average $S_{th}$ decreasing from 2.7 to 2.0 V dec$^{-1}$. I$_2$ concentration of higher than 0.5 mol% led to a decrease in $\mu_{sat}$, a negative shift in $V_{th}$, and an increase in $S_{th}$ although the doped devices still showed superior performance to the pristine devices up to ~10 mol%. The improvement in the hole transport properties is further evident from the power law analysis (see Supporting Information) of the linear hole mobility ($\mu_{lin}$) as shown in Figure 4d.[68] The reduction in the exponent $\gamma$ from 0.71 to 0.13 when doped with 0.5 mol% I$_2$ indicates that the hole transport mechanism in CuSCN changes from the slower trap-limited conduction (TLC) to the more mobile percolation conduction (PC).[30,63]

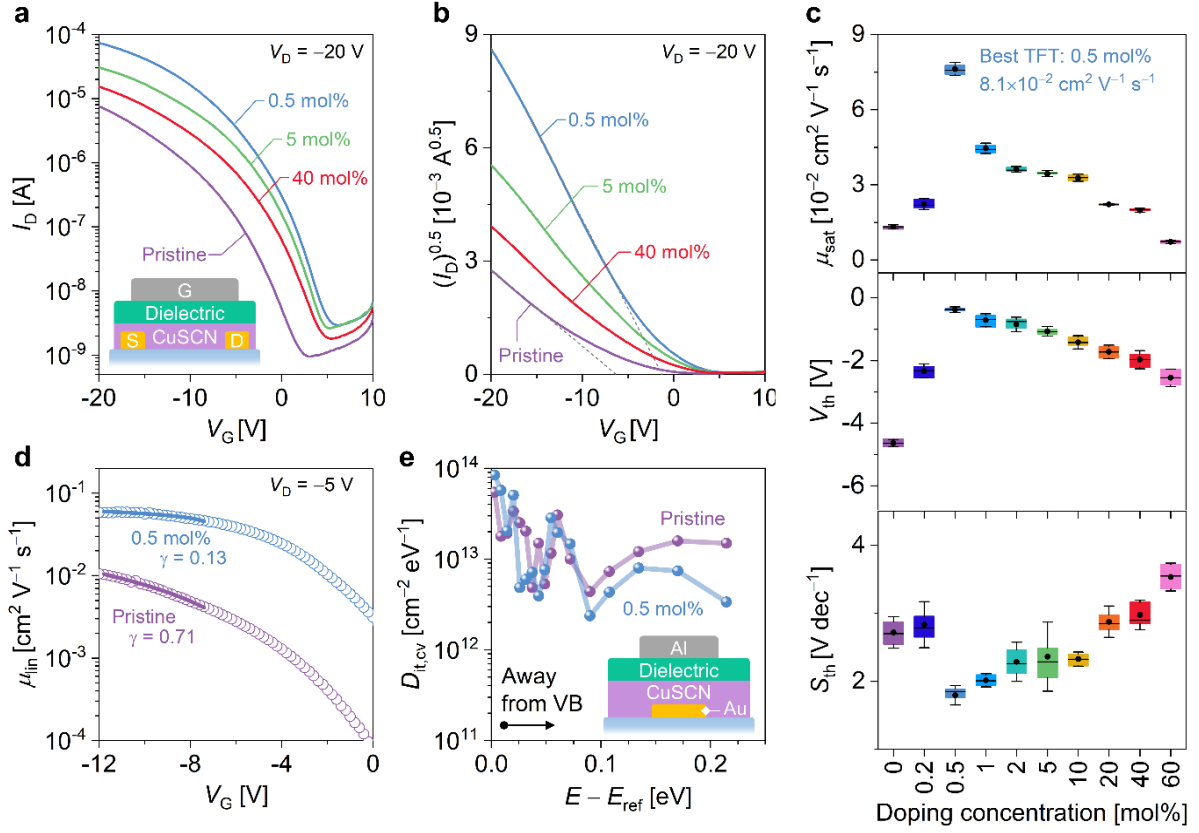

**Figure 4.** Transfer characteristics (forward sweep) in the saturation regime plotted as (a) log($I_D$) vs $V_G$ and (b) $\sqrt{I_D}$ vs $V_G$. (a, inset) Schematic diagram of the top-gate bottom-contact thin-film transistors. Device dimensions: $W$ = 1000 μm and $L$ = 30 μm. (c) Box plots of $\mu_{sat}$, $V_{th}$, and $S_{th}$. The box and the horizontal line represent the 25$^{th}$, 50$^{th}$, and 75$^{th}$ percentiles. The black dots show the mean values, and the whiskers the standard deviation. (d) Power law analysis of $\mu_{lin}$. (e) Results of trap state density analysis from the frequency dispersion of the metal-insulator-semiconductor capacitors (structure shown in the inset).

The positive shift in $V_{th}$ also signifies an increase in the hole density ($N_h$) in the channel as shown in Figure S12a; in the I$_2$-doped CuSCN channel, a smaller number of electric field-induced holes is required to turn the device on (or equivalently, at the same applied $V_G$, there is a higher hole density in the I$_2$-doped CuSCN channel). The reason for this could be from:

(1) a p-doping effect, i.e., an increase in the hole density in the semiconductor layer; or (2) a decrease in the hole-trapping states.[38,69] The p-doping effect in CuSCN can be clearly seen from the work by Wijeyasinghe et al., in which a strong electron acceptor $C_{60}F_{48}$ was used as a dopant.[30] In their study, the positive shift in $V_{th}$ was accompanied by a large increase in $I_{D,off}$ which led to a reduction in the on/off ratio from $2\times10^3$ (undoped CuSCN) to $1\times10^3$ (0.5 mol% $C_{60}F_{48}$) despite the gain in $I_{D,on}$. The p-doping was also manifested in the increase in $S_{th}$; the added dopants gave rise to more impurity scattering.[70] A similar p-doing behavior in CuSCN TFTs can also be observed from the report by Ji et al.[35] In contrast, the p-doping effect was negligible in our results. $I_{D,off}$ was only slightly increased from 1 nA to 3 nA upon doping whereas the on/off ratio increased from $8\times10^3$ (undoped CuSCN) to $2\times10^4$ (0.5 mol% $I_2$) due to the large increase in $I_{D,on}$. Furthermore, $S_{th}$ was in fact reduced. This evidence strongly highlights that the addition of $I_2$ leads to a reduction in the density of hole-trapping states. For an estimation, the interface trap density calculated from $S_{th}$ of TFTs ($D_{it,tft}$) decreased from $6.1\times10^{13}$ to $4.5\times10^{13}$ $cm^{-2}$ $eV^{-1}$ at 0.5 mol% $I_2$ concentration as displayed in Figure S12b.

**6. Trap States Passivation, Contact Resistance, and Bias Stability**

To characterize the trap states in more detail, we studied the frequency dispersion of the capacitance-voltage (*C-V*) characteristics of metal-insulator-semiconductor (MIS) devices[71,72] for the pristine and 0.5 mol% $I_2$-doped CuSCN. The MIS structure was fabricated using the same layer sequence as the TFT structure as illustrated in the inset of Figure 4e. The resulting *C-V* characteristics measured at various frequencies are shown in Figure S13a and b. Both devices showed the typical hole-accumulation behavior when the voltage was swept to negative values. The onset voltages ($V_{onset}$) of the hole accumulation were shifted to more positive values with the decreasing frequency as shown in Figure S13c. At high frequencies, only trap states near the valence band edge are responsive; as the frequency decreases, more trap states at energies deeper in the tail start to respond, leading to higher capacitance and the shift in $V_{onset}$.[72] The latter allows the analysis of the energetic distribution of the interface trap density ($D_{it,cv}$) as shown in Figure 4e (see Supporting Information for detail). A rapid change in $V_{onset}$ is associated with a large trap density; in contrast, a gradual change in $V_{onset}$ is due to a small trap density. In the energy range of 0.0-0.1 eV with respect to the reference energy (corresponding to the highest frequency measured), the $D_{it,cv}$ profiles of both pristine and $I_2$-doped CuSCN samples were similar. However, the broad feature in the 0.1-0.2 eV range seen in the pristine sample became significantly smaller upon the addition of 0.5 mol% $I_2$, showing the decrease in trap states upon doping.

The reduction in the trap state density, evident from both $D_{it,tft}$ (based on TFT subthreshold characteristics) and $D_{it,cv}$ (based on the frequency dispersion of MIS C-V characteristics), was concurrent with the defect healing effect and improved crystallinity by $I_2$ doping as discussed in the preceding sections. We surmise that the pristine CuSCN films prepared from the DES solution contain a large number of hole traps from $V_{SCN}$ (high $D_{it,tft}$ and $D_{it,cv}$ with under-coordinated $Cu^+$ centers). Indeed, excess SCN is known to increase the p-type conductivity of CuSCN;[25,73] accordingly, the deficiency in SCN (i.e., the formation of $V_{SCN}$) is then expected to reduce the hole concentration. This is also supported by a theoretical study of defects in CuSCN which reported that $V_{SCN}$ form donor states within the band gap of CuSCN.[27] Electron-donating states are known to trap holes and result in the negative shift in $V_{th}$,[74] which is consistent with our results of the pristine CuSCN. At the optimal condition of $I_2$ addition (0.5 mol%), the incorporated iodine exists as $I^-$ which fills in the void of $V_{SCN}$ and effectively suppresses the hole-trapping states (low $D_{it,tft}$ and $D_{it,cv}$ with fully coordinated $Cu^+$). A recent study also reported on a similar passivation of iodine vacancies by $Br^-$ and $Cl^-$ in tin halide perovskite TFTs.[75]

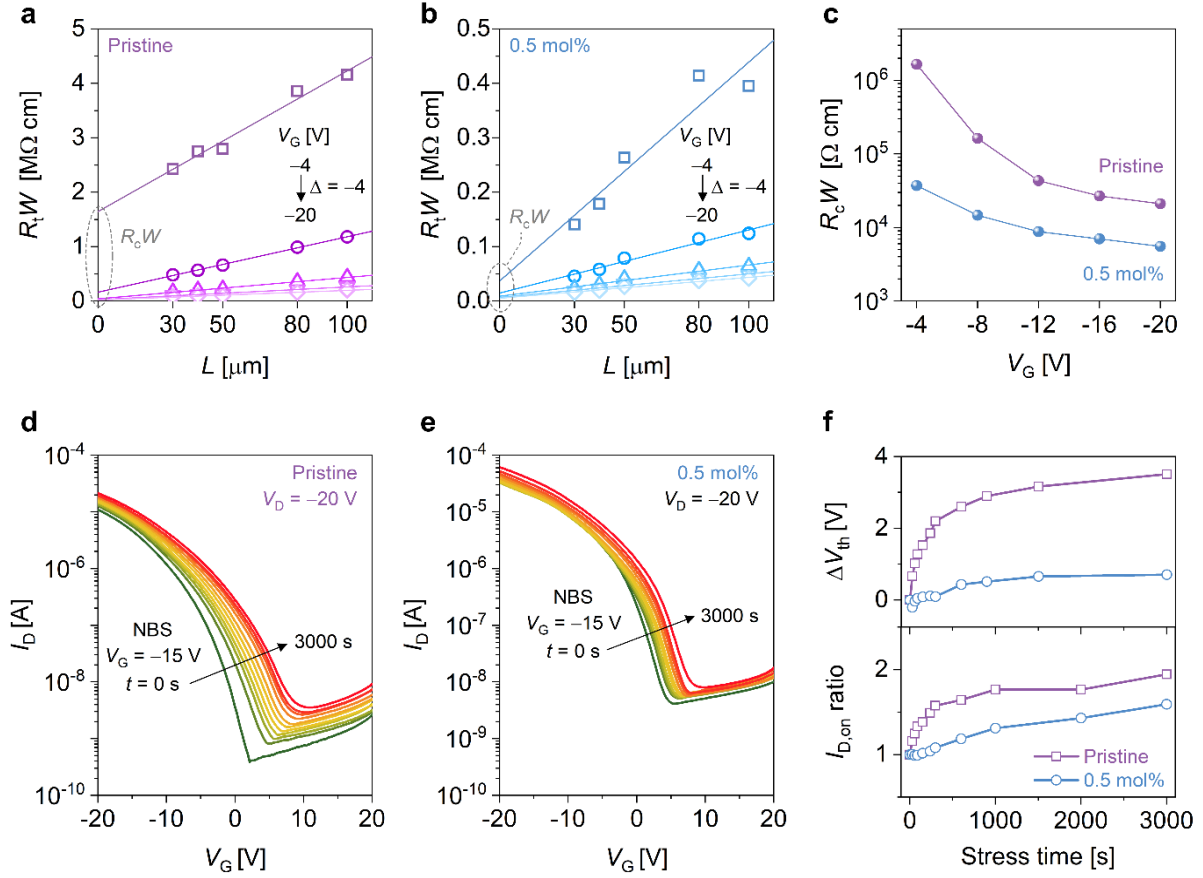

**Figure 5.** Width-normalized total resistance ($R_tW$) of TFTs based on (a) pristine and (b) 0.5 mol% $I_2$-doped CuSCN layers vs channel length ($L$). The width-normalized contact resistance ($R_cW$) can be calculated from the $y$-intercept of the linear fit. (c) $R_cW$ vs $V_G$. (d), (e) Transfer characteristics during the negative bias stress (NBS) stability test of TFTs based on pristine and 0.5 mol% $I_2$-doped CuSCN layers, respectively. (f) The shift in the threshold voltage ($\Delta V_{th}$) and the change in $I_{D,on}$ (given as a ratio to the value at $t = 0$ s) from the bias-stress stability test.

We also analyzed the contact resistance ($R_c$) in CuSCN TFTs using the transmission line method (see Supporting Information). **Figure 5**a and b show the width-normalized total resistance ($R_tW$) of pristine and $I_2$-doped CuSCN TFTs as a function of the channel length ($L$) at varying $V_G$. The width-normalized contact resistance ($R_cW$) was determined from the $y$-intercept, and the results are plotted in Figure 5c. The addition of 0.5 mol% $I_2$ drastically reduced $R_cW$: from 1.6 MΩ cm to 37 kΩ cm at $V_G$ = -4 V or from 21 kΩ cm to 5.6 kΩ cm at $V_G$ = -20 V. The large decrease in the contact resistance can be attributed to the effects of defect healing and improved hole transport properties in the semiconductor layer.[76,77]

In addition, bias-stress stability measurements were performed to demonstrate further the advantages of trap passivation in CuSCN TFTs. A negative bias stress (NBS) of $V_G$ = -15 V was applied to the devices based on pristine and 0.5 mol% $I_2$-doped CuSCN layers. Figure 5d and e display the transfer characteristics in the saturation regime ($V_D$ = -20V) measured at different time intervals during a 3000-s period. It is clear that the $I_2$ doping alleviates the adverse effects of NBS; the changes in $V_{th}$ and $I_{D,on}$ were effectively reduced as shown in Figure 5f. The changes in the key TFT parameters calculated from the transfer dataset are listed in Table S7. Interestingly, the NBS led to a positive $V_{th}$ shift by +3.5 V for the device based on pristine CuSCN, concomitantly with increases in $I_{D,on}$ by almost two times, $I_{D,off}$ by nine times, and $S_{th}$ by +1.6 V dec$^{-1}$. This signifies a creation of acceptor states under the NBS condition that led to an apparent p-doping effect. Although the identity of these NBS-induced acceptor states is unknown at this point, we speculate that they may be related to Cu centers, e.g., Cu vacancies. Our rationale is based on the observation that the NBS-induced p-doping effect was also suppressed in the case of 0.5 mol% $I_2$-doped CuSCN. For the latter, $V_{th}$ was also shifted positively but only by 0.7 V whereas $I_{D,on}$ was increased by 60%, $I_{D,off}$ by about two times, and $S_{th}$ by 0.2 V dec$^{-1}$. The more complete coordination around Cu$^+$ in the optimally doped CuSCN may give rise to more robustness of the Cu centers against the effects of the NBS.

## 7. Conclusion

We have shown that SCN vacancies are a major source of defect states in CuSCN processed from the widely used DES-based recipe. These defects act as hole-trapping states that restrict the hole transport properties of pristine CuSCN films. Though a simple method of adding $I_2$ to the processing solution, the properties of CuSCN can be remarkably improved. At small concentrations, the added iodine exists as $I^-$ which can replace $V_{SCN}$ and effectively restore the coordination environment of $Cu^+$. This defect-healing effect results in drastically enhanced hole transport properties, evident from the improvements of TFT parameters across the board at the optimal doping concentration of 0.5 mol%. Specifically, the hole mobility was increased from around 0.01-0.02 to 0.07-0.08 cm$^2$ V$^{-1}$ s$^{-1}$ while the on/off current ratio was simultaneously increased from 8×10$^3$ to 2×10$^4$. The latter is distinct from previous doping attempts; the optimal $I_2$ doping can passivate hole traps while avoiding the p-doping effect, hence preventing the undesired increase in $I_{D,off}$ and the reduction in the on/off current ratio. Furthermore, the contact resistance is reduced, and the device stability under NBS is more robust. As the $I_2$ doping introduced herein is extremely simple and based on a cost-effective, commercially available chemical, the process could be extended to a wide array of devices that already employ CuSCN as the HTL, such as PSCs, OSCs, and OLEDs.

In addition, this work highlights that there is still a large room for CuSCN property optimization. Although $I^-$ is shown to recover the coordination environment around $Cu^+$ and passivate the hole-trapping states, the trap state density is still relatively high (on the order of 10$^{13}$ cm$^2$ V$^{-1}$ s$^{-1}$) compared to high-performance p-channel TFTs based on other semiconductor families, such as organics or oxides (with trap state density on the order of 10$^{11}$-10$^{12}$ cm$^2$ V$^{-1}$ s$^{-1}$).[78–80] We therefore expect that further defect engineering will be crucial to advance the performance of CuSCN-based devices beyond the current state of the art.

**Experimental Section**

Full descriptions of the experiments are given in the Supporting Information.

**Acknowledgements**

P.W., S.W., and P.P. would like to acknowledge funding from Vidyasirimedhi Institute of Science and Technology (VISTEC) and support of scientific instruments from VISTEC's Frontier Research Center (FRC). This project is funded by National Research Council of Thailand (NRCT), grant no. N42A650254 and N42A650196, and Thailand Science Research and Innovation (TSRI), grant no. FRB650023/0457.

**Conflict of Interest**

The authors declare no conflict of interest.

# Origin of Hole-Trapping States in Solution-Processed Copper(I) Thiocyanate (CuSCN) and Defect-Healing by I$_2$ Doping

Pimpisut Worakajit,[1] Pinit Kidkhunthod,[2] Saran Waiprasoet,[1] Hideki Nakajima,[2] Taweesak Sudyoadsuk,[1] Vinich Promarak,[1] and Pichaya Pattanasattayavong[1,*]

[1] Department of Materials Science and Engineering, School of Molecular Science and Engineering, Vidyasirimedhi Institute of Science and Technology (VISTEC), Rayong 21210, Thailand

[2] Synchrotron Light Research Institute (Public Organization), 111 University Avenue, Muang, Nakhon Ratchasima, 30000, Thailand

* Correspondence: pichaya.p@vistec.ac.th

**Experimental Section**

***Solution Preparation.*** CuSCN and I$_2$ stock solutions were prepared by separately dissolving copper(I) thiocyanate powder (CuSCN, 99%, Sigma-Aldrich) and iodine beads (I$_2$, 99.9%, Wako) in diethyl sulfide (DES, 98%, TCI) at a concentration of 10 mg ml$^{-1}$. Both solutions were stirred overnight and filtered using a 0.22-μm nylon filter in a N$_2$-filled glove box. To prepare mixed CuSCN-I$_2$ solutions, the stock I$_2$ solution was added to the stock CuSCN solution with various mixing volumes as described in Table S1. The mixed solutions were continuously stirred for at least 30 min at room temperature before usage.

***Thin-film Fabrication.*** Substrates were cleaned using a sequential ultrasonication procedure in 1%v/v detergent solution (Liquinox, Alconox Inc.), deionized water, acetone, and isopropanol, each for 30 min. Prior to use, the substrates were dried using N$_2$ gas and treated with UV-ozone for wettability improvement. CuSCN and mixed CuSCN-I$_2$ solutions were deposited onto the substrates by spin-coating at 2000 rpm for 60 s. The resultant films were annealed at 100 °C for 30 min. The spin-coating and annealing steps were performed in a N$_2$-filled glove box.

***Drop-cast Sample Preparation.*** For certain characterization techniques, i.e., Raman spectroscopy, X-ray diffraction, and X-ray absorption spectroscopy, drop-cast samples of CuSCN and I$_2$-doped CuSCN were employed for the measurements to improve the signal intensity and result quality. The solutions were drop-cast onto substrates which had been pre-heated to 100 °C and then kept at that temperature for 30 min to evaporate the solvent. The remaining solid samples on the substrates were collected and stored inside the glove box before the characterizations.

***Raman Spectroscopy.*** Raman spectra of CuSCN and $I_2$-doped CuSCN (prepared by drop-casting) were recorded using a Bruker Senterra Raman microscope with a dispersive Raman spectrometer. The excitation source was a 532-nm laser operated at 2 mW. All spectra were collected using a 10x objective lens and 20 s of acquisition time.

***X-ray Diffraction (XRD).*** The powder samples were prepared using the abovementioned drop-casting procedure. XRD spectra were recorded with a Bruker D8 Advance using a Cu $K_\alpha$ X-ray source ($\lambda$ = 1.5406 Å). The spectra were collected in the 2$\theta$ range of 5° to 60° under ambient environment.

***X-ray Absorption Spectroscopy (XAS).*** Similar samples to those used for Raman and XRD measurements were also employed for XAS. The samples were manually ground before packing in Kapton sleeves for measurements. XAS was conducted at beamline BL5.2 SUT-NANOTEC-SLRI, Synchrotron Light Research Institute (SLRI), Nakhon Ratchasima, Thailand (electron energy 1.2 GeV, bending magnet, beam current 80-150 mA). The standard samples were CuSCN powder (99%, Sigma-Aldrich), CuI powder (99.5%, Sigma-Aldrich), and $I_2$ beads (99.9%, Wako). The XAS measurements of both Cu K-edge and I $L_3$-edge, covering both the near edge structure (XANES) and the extended fine structure (EXAFS), were conducted in fluorescent mode under ambient conditions. The spectra were averaged from five scans and analyzed by the standard procedure using the Demeter package (ATHENA and ARTEMIS)[1]. The spectra analysis processes included background subtraction, normalization, and $\chi$(k) isolation. The EXAFS signals were extracted by Fourier transform using the Hanning function.

***Film Morphology and Thickness Characterizations.*** Surface topography of pristine and $I_2$-doped CuSCN films were characterized by a Park Systems NX10 atomic force microscope (AFM) operated in the non-contact mode with an Olympus OMCL-AC160TS cantilever (Si, 7-nm tip radius, 26-N m$^{-1}$ stiffness, 300-kHz resonance frequency). Surface height ($Z_{AFM}$) histograms and root-mean-square roughness ($\sigma_{RMS}$) values were obtained from 5×5 µm$^2$ images while average grain diameter values ($D_{avg}$) were calculated from 0.5×0.5 µm$^2$ images. Gwyddion software (http://gwyddion.net/) was used to analyze the AFM images.[2] The thicknesses of the films were measured with a Bruker Dektak XT profilometer. For each condition, two samples, each measured on four different spots, were used to find the average thickness. The error was given by the standard deviation.

***Ultraviolet-visible-near-infrared Spectroscopy (UV-Vis-NIR).*** CuSCN and $I_2$-doped CuSCN thin-film samples were prepared using the method described above onto fused silica substrates (to minimize absorption in the UV range from the substrates). Optical transmittance and reflectance spectra were recorded in the range of 200-1400 nm using a PerkinElmer LAMBDA 1050 spectrophotometer equipped with an integrating sphere. The absorbance spectra $\alpha(\lambda)$ were calculated by considering the film-substrate interface effect using the equation:[3]

$$\alpha(\lambda) = \ln\left(\frac{1-R_{tot}}{T_{tot}}\right) - \ln\left(\frac{1-R_{sub}}{T_{sub}}\right) \tag{S1}$$

where $R_{tot}$ and $R_{sub}$ are the total reflectance of the sample and reflectance of the bare substrate, respectively, and $T_{tot}$ and $T_{sub}$ are the total transmittance of the sample and transmittance of the bare substrate, respectively. The optical band gaps were determined using Tauc plots.

***Photoelectron Spectroscopy (PES).*** The secondary electron cutoff and valence band spectra were recorded at Beamline BL3.2Ua, Synchrotron Light Research Institute, Nakhon Ratchasima, Thailand, using a photon energy 60 eV. The samples were prepared onto ITO substrates in a $N_2$-filled glove box before transferred to the vacuum chamber (base pressure of $10^{-8}$ Pa). Au was used as a reference sample and all samples were biased at about 9.5 V to allow the measurements of low energy electrons at the secondary electron cutoff edge. The results were recorded with pass energy of 10 eV using a Thermo VG Scientific CLAM2 electron energy analyzer, yielding a total energy resolution of about 0.2 eV. The work functions of the samples were determined from the secondary electron cutoff edges. After aligning the Fermi level at 0 eV, the valence band edges were determined from the valence band onsets. The conduction band edges were estimated by adding the optical band gaps (from UV-Vis-NIR) to the valence band positions.

***Thin-film Transistor (TFT) Fabrication and Characterization.*** TFT devices were fabricated in the top-gate bottom-contact (TG-BC) structure. First, 10-nm Cr/40-nm Au source-drain (S-D) electrodes were sequentially deposited onto 1×1 in$^2$ borosilicate glass substrates by thermal evaporation using a KJ Lesker Minispectros system. Then, CuSCN and $I_2$-doped CuSCN films were deposited following the spin-coating procedure as mentioned above. Next, a dielectric layer, prepared from a solution of poly(vinylidene fluoride-trifluoroethylene-chlorofluoroethylene) or P(VDF-TrFE-CFE) (63/30/7 mol%, PolyK Technologies) dissolved in 2-butanone (>99%, TCI) at 25 mg ml$^{-1}$, was spin-cast on the semiconductor layer at 3000 rpm for 30 s and annealed at 80 °C for 3 h. The thickness and the geometric capacitance of the dielectric layer was 200 nm and 211.8 nF cm$^{-2}$ as measured by profilometry and impedance spectroscopy, respectively. Finally, 40-nm Al was thermally evaporated on top of the dielectric layer as a gate electrode to complete the transistor structure. Both Au and Al electrodes were deposited through a shadow mask, resulting in TFT devices with a channel width ($W$) of 1000 µm and a channel length ($L$) of 30 µm. The electrical characterizations of the TFT devices were carried out with a probe station under $N_2$ atmosphere using a Keysight B2912A 2-channel source-measure unit (SMU). TFT device parameters, i.e., linear and saturation mobility ($\mu_{lin}$ and $\mu_{sat}$), threshold voltage ($V_{th}$), and subthreshold swing ($S_{th}$), were calculated using the standard gradual channel approximation.[4] The change in the hole density ($\Delta N_h$) of the devices was estimated based on the shift in the threshold voltage ($\Delta V_{th}$) via[5]

$$\Delta N_h = \frac{C_i \Delta V_{th}}{q} \tag{S2}$$

where $C_i$ is the geometric capacitance of the dielectric layer. The interface trap density $D_{it,tft}$ (expressed in eV$^{-1}$ cm$^{-2}$) was evaluated from $S_{th}$ using the equation:[6,7]

$$D_{it,tft} = \frac{C_i}{q}\left(\frac{q}{kT}\cdot\frac{S_{th}}{\ln 10}-1\right) \tag{S3}$$

The charge carrier transport mechanisms were analyzed based on the power law:[8]

$$\mu_{lin} = K\left(V_G - V_{th}\right)^{\gamma} \tag{S4}$$

where $K$ is the proportional constant, $V_G$ the gate voltage, and $\gamma$ the fitting exponent. A large exponent $\gamma$ indicates that the conduction mechanism is dominated by the trap-limited conduction (TLC) in which carriers are trapped and released multiple times between the extended states (related to band states) and the localized tail states. In contrast, a small $\gamma$ is associated with the percolation conduction (PC), which refers to the mechanism when charge carriers find the path with the least resistance (blocked by randomly distributed potential barriers) to move across the extended states.[8]

***Capacitance-voltage (C-V) Characterization.*** Metal-insulator-semiconductor (MIS) devices were fabricated on borosilicate glass substrates using a similar layer stacking to the TFT devices. The bottom electrode was 10-nm Cr/25-nm Au, and the top electrode was 100-nm Al. The cross pattern of the two electrodes yielded a device area of 0.04 cm$^2$. Pristine or I$_2$-doped CuSCN films and P(VDF-TrFE-CFE) were deposited in between the bottom and top electrodes following the procedure described above to form the semiconductor and the insulator layer, respectively. The final structure was Au/CuSCN/P(VDF-TrFE-CFE)/Al. The *C-V* characteristics were measured with different AC frequencies using a Solartron Analytical 1260A impedance analyzer. All fabrication and characterization steps were carried out in a N$_2$-filled glove box.

From the *C-V* characteristics, the interface trap density $D_{it,cv}$ (in eV$^{-1}$ cm$^{-2}$) can be estimated for an energy range between $E_j$ and $E_{j+1}$ (where j = 1, 2, 3, …) from the shift in the onset voltage ($V_j$ and $V_{j+1}$) due to the frequency dispersion of the *C-V* curves (measured at different frequencies $f_j$ and $f_{j+1}$, where $f_j > f_{j+1}$) using the equation:[9,10]

$$D_{it,cv} = \frac{C_i}{q}\cdot\frac{|V_{j+1}-V_j|}{E_{j+1}-E_j} = \frac{C_i|V_{j+1}-V_j|}{kT\ln\left(f_j/f_{j+1}\right)}, \tag{S5}$$

where $C_i$ is the geometric capacitance of the dielectric layer.

In short, the principle of the method is that the interface trap states closer to the mobile states (near the valence band edge in this case), which have faster time constants, can respond to high-frequency signals whereas states deeper in the gap (away from the valence band), which have slower time constants, cannot. The latter then respond to lower frequencies.[10] By comparing a *C-V* curve measured at a higher frequency ($f_j$) to that measured at a lower frequency ($f_{j+1}$), the capacitance of the latter should be higher due to more interface trap states becoming more responsive; this is accompanied by a shift in the onset voltage from $V_j$ (measured at $f_j$) to $V_{j+1}$ (measured at $f_{j+1}$). The number density ($N_{it,cv}$) of the interface trap states can therefore be estimated from: $N_{it,cv} = C_i\cdot|V_{j+1} - V_j|/q$. The corresponding energy range $\Delta E = E_{j+1} - E_j$ (where $E_{j+1}$ is farther away from the valence band than

$E_j$, corresponding to $f_{j+1}$ and $f_j$, respectively) is calculated (in eV) from: $E_{j+1} - E_j = (kT/q)\cdot\ln(f_j/f_{j+1})$. The average density of states ($D_{it,cv}$) is therefore: $D_{it,cv} = N_{it,cv}/\Delta E$, which is Eq. S5. The resulting $D_{it,cv}$ is plotted at an average energy ($E_{avg}$) between $E_j$ and $E_{j+1}$: $E_{avg} = E_j + (\Delta E/2)$. By comparing pairs of C-V curves measured at successively lower frequencies ($f_1$ vs $f_2$, $f_2$ vs $f_3$, and so on, where $f_1 > f_2 > f_3 > \ldots$), a spectrum of $D_{it,cv}$ can be constructed. The frequencies employed for our measurements and the resulting onset voltages are listed in Table S6.

$E_1$ can be calculated using the Shockley-Read-Hall statistics.[10] However, in our case, many parameters required for the calculation are still unknown, i.e., the capture cross section of the interface trap states, effective density of states of the band, and thermal velocity of electrons. Therefore, we plot the $D_{it,cv}$ spectra vs the energy distance from $E_1$, taken as a reference level ($E_{ref}$) which is assumed to be close to the valence band edge.

***Contact Resistance Analysis.*** The contact resistance ($R_c$) of the devices was determined by the transmission line method following the equation:[11]

$$R_t = \frac{\partial V_{D,lin}}{\partial I_{D,lin}} = R_{ch} + R_c = R_s \frac{L}{W} + R_c \tag{S6}$$

$$R_t W = R_s L + R_c W \tag{S7}$$

where $R_t$ is the total device resistance which can be found from the output characteristics in the linear region ($I_{D,lin}$ vs $V_{D,lin}$). $R_t$ has two components: contact resistance ($R_c$) which in this case combines the contribution from both the source and drain contacts; and channel resistance ($R_{ch}$) which can be written in terms of semiconductor resistance ($R_s$) and device dimensions ($L$ and $W$). By measuring $R_t$ of devices with different $L$, a linear extrapolation to the limit $L = 0$ allows the determination of $R_c$. The normalization by the channel width ($W$), as in Eq. S7, allows for a comparison across devices with different geometry.

***Bias Stability Measurement.*** The stability under gate-bias stress of TFTs based on pristine and $I_2$-doped CuSCN was measured by applying a negative $V_G$ bias of –20 V (using the same setup for the TFT measurements) under a $N_2$-filled glove box. Transfer characteristics were collected at different times during the stressing period of 0 to 3000 s.

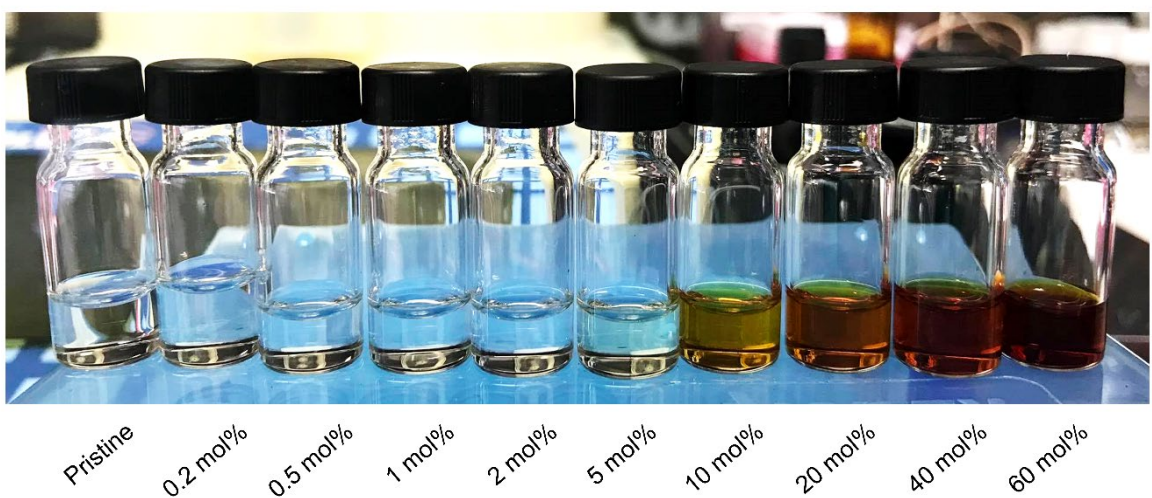

**Figure S1.** Photographs of the vials containing CuSCN solutions with different $I_2$ doping concentrations.

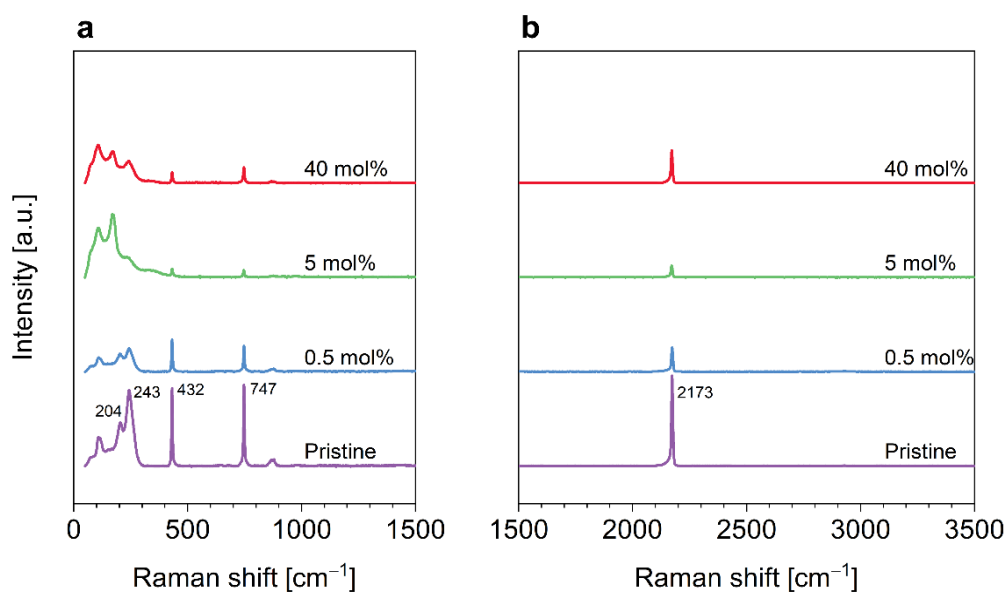

**Figure S2.** Raman spectra of pristine and $I_2$-doped CuSCN samples (drop-cast from DES solutions) in the (a) 0 to 1500 cm$^{-1}$ and (b) 1500 to 3500 cm$^{-1}$ wavenumber ranges.

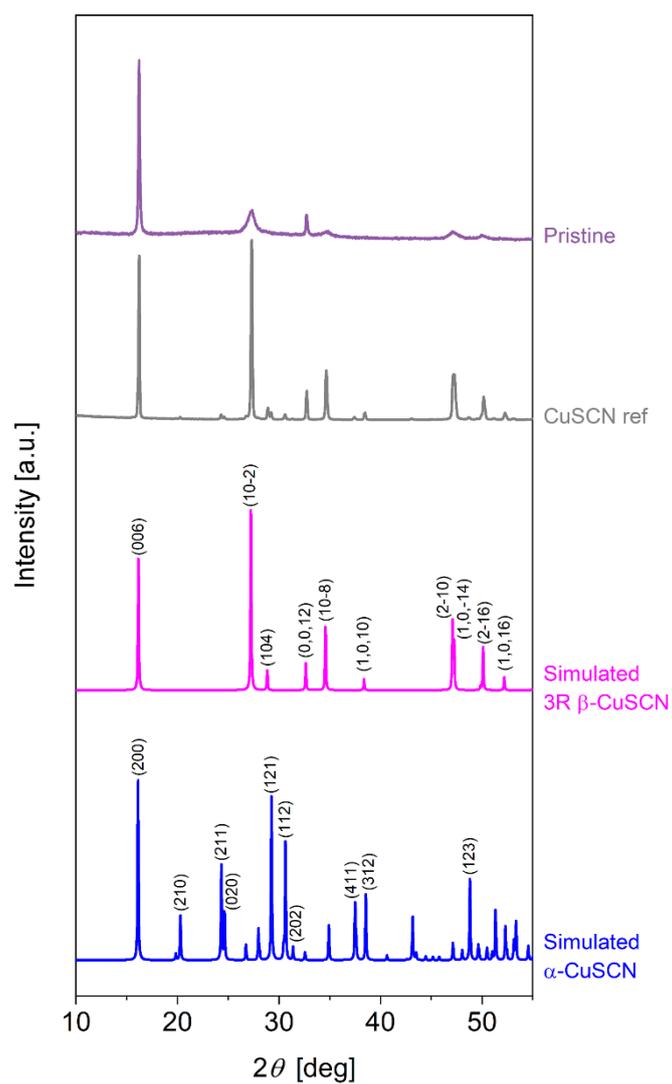

**Figure S3.** Comparison of the X-ray diffraction pattern of the prisine CuSCN sample (drop-cast from DES solution) with that of the reference CuSCN powder (99%, Sigma-Aldrich), simulated pattern of 3R β-CuSCN (ICSD 24372), and simulated pattern of α-CuSCN (ICSD 124).

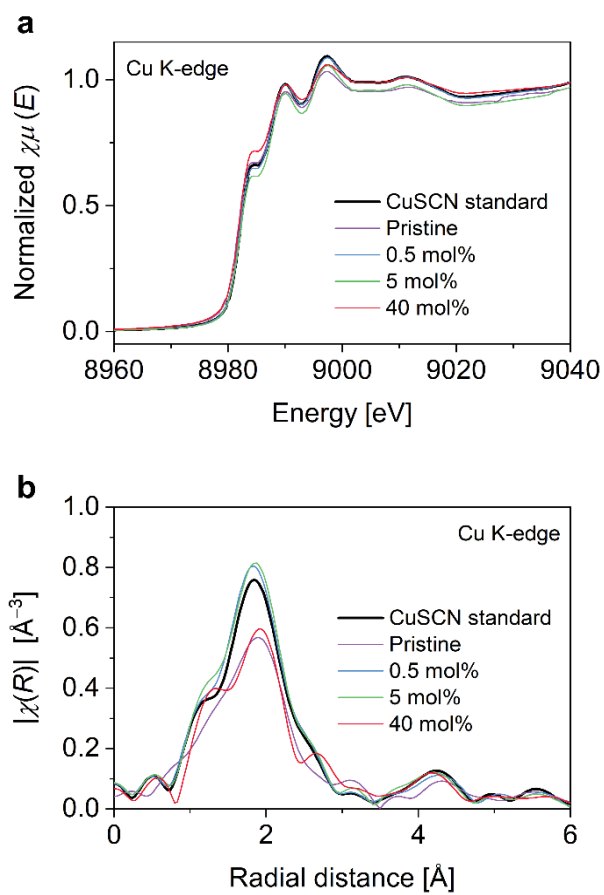

**Figure S4.** (a) XANES spectra at the Cu K-edge. (b) Analyzed EXAFS spectra in the R-space.

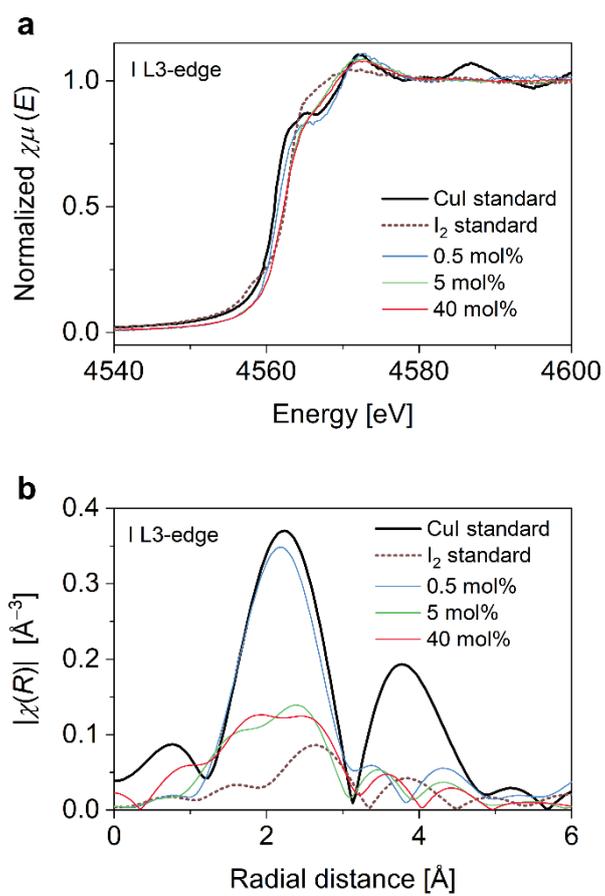

**Figure S5.** (a) XANES spectra at the I L3-edge. (c) Analyzed EXAFS spectra in the R-space.

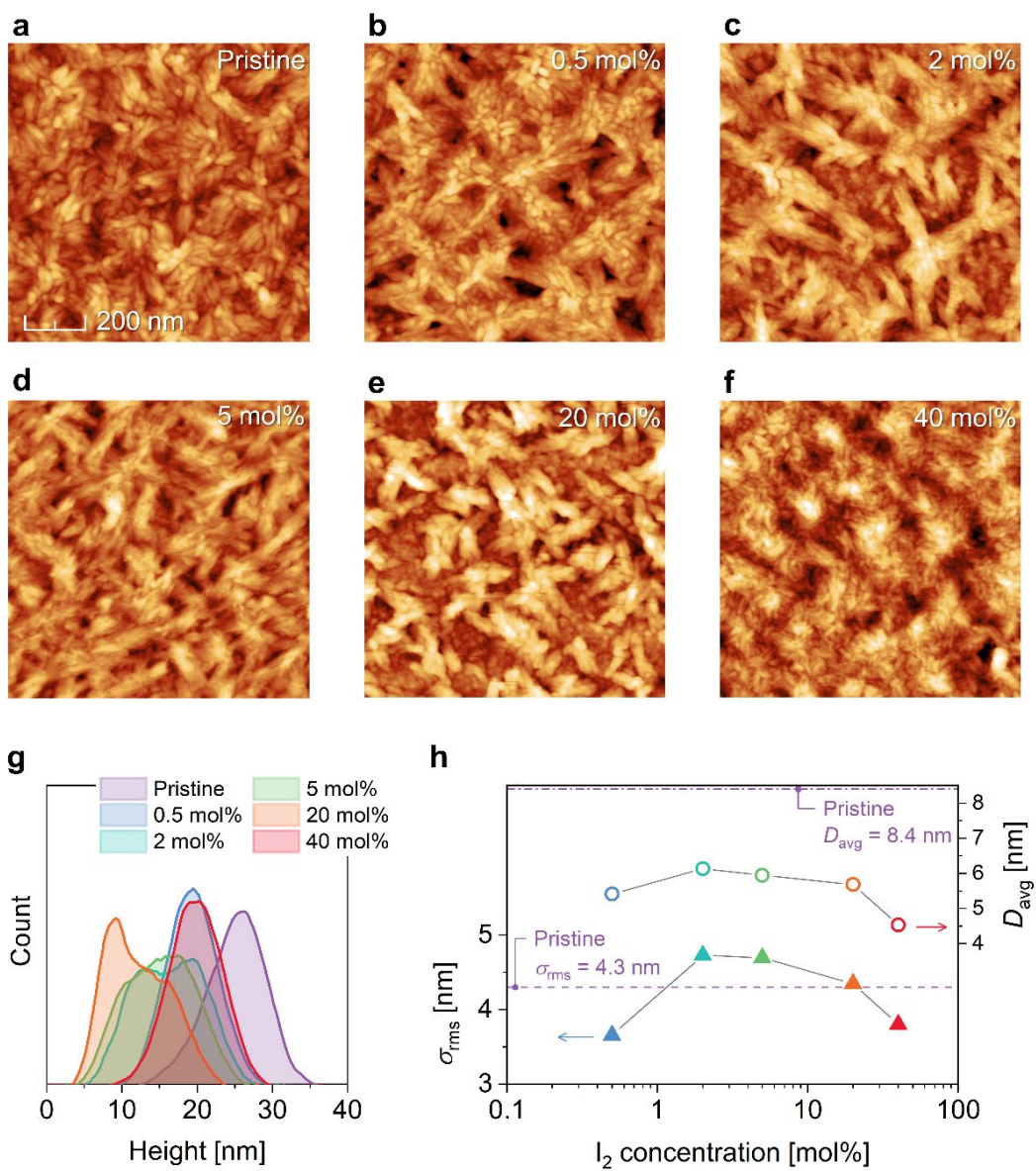

**Figure S6.** (a-f) AFM images of pristine and I$_2$-doped CuSCN films of area of 1x1 µm$^2$. Analyzed data: (g) surface height histograms and (h) root-mean-squre surface roughness ($\sigma_{rms}$, left y-axis, solid triangles) and average grain size ($D_{avg}$, right y-axis, open circles) of various I$_2$ doping conditions. The dashed lines represent data of the pristine CuSCN films: $\sigma_{rms}$ = 4.3 nm and $D_{avg}$ = 8.3 nm.

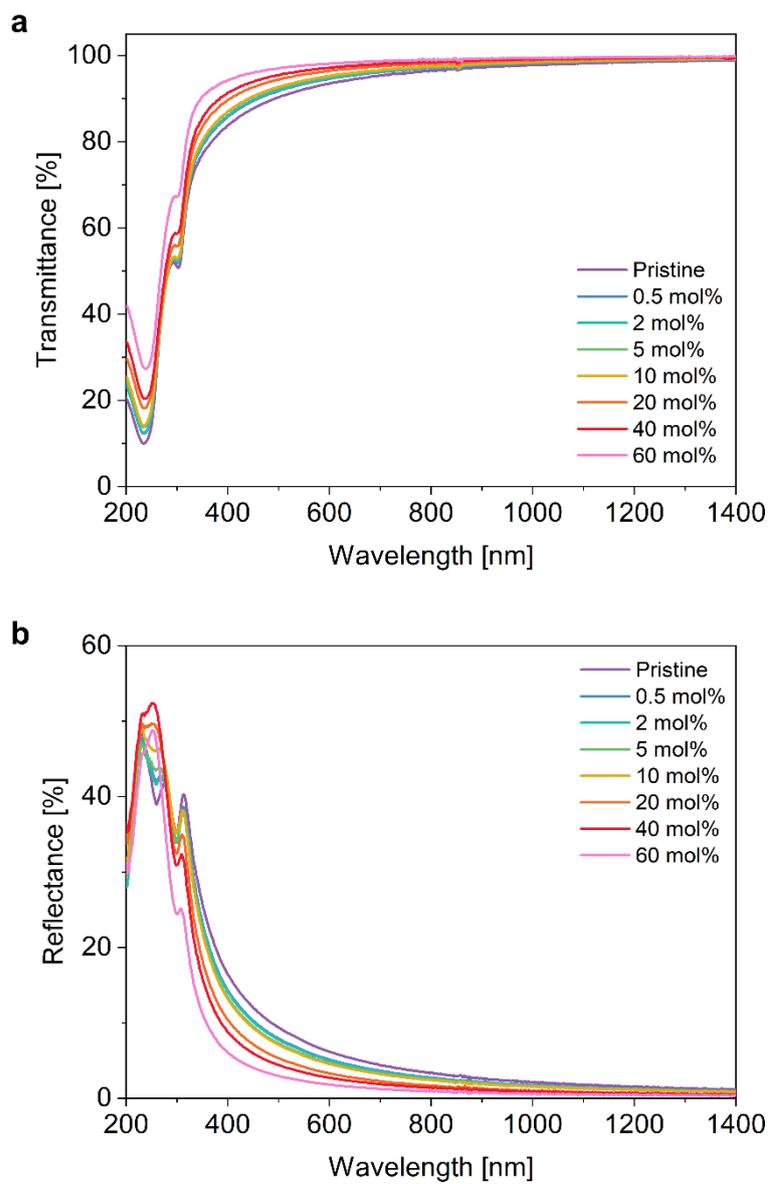

**Figure S7.** UV-Vis-NIR (a) transmittance and (b) reflectance spectra of pristine and $I_2$-doped CuSCN thin films spin-cast on fused silica substrates.

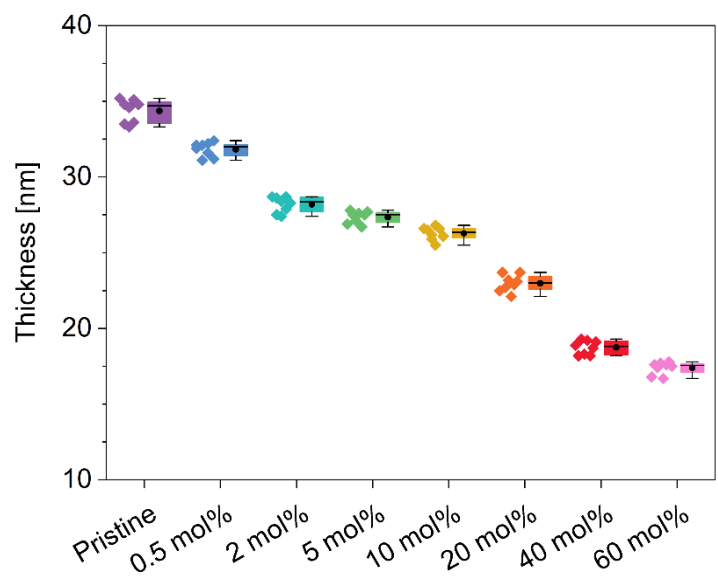

**Figure S8.** Thickness of pristine and $I_2$-doped CuSCN thin films measured from profilometry.

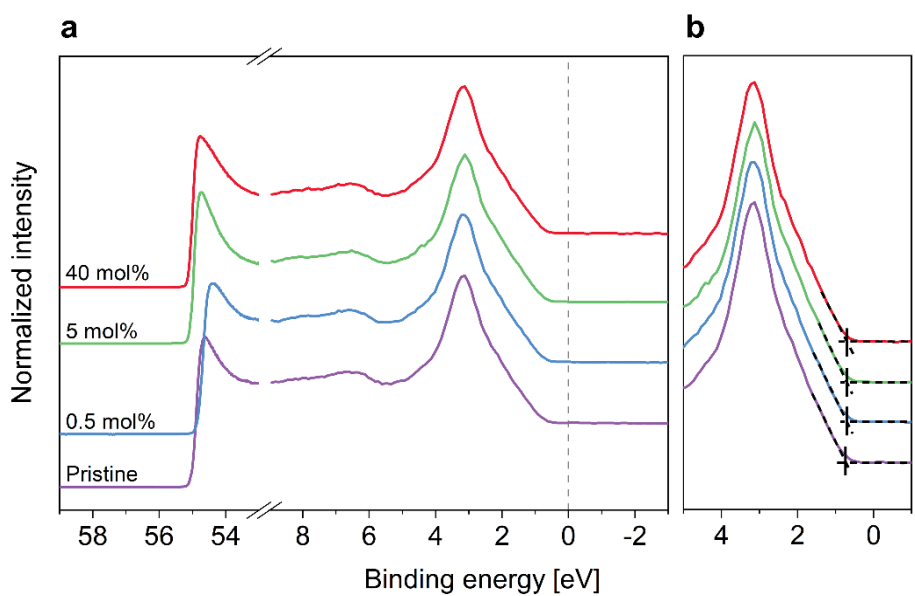

**Figure S9.** Photoemission spectra of pristine and $I_2$-doped CuSCN films on ITO substrates measured using synchrotron radiation. (a) Full data range showing the top of the valence bands and the secondary electron cutoff edges. (b) Close-up at the top of the valence bands for the determination of the onsets.

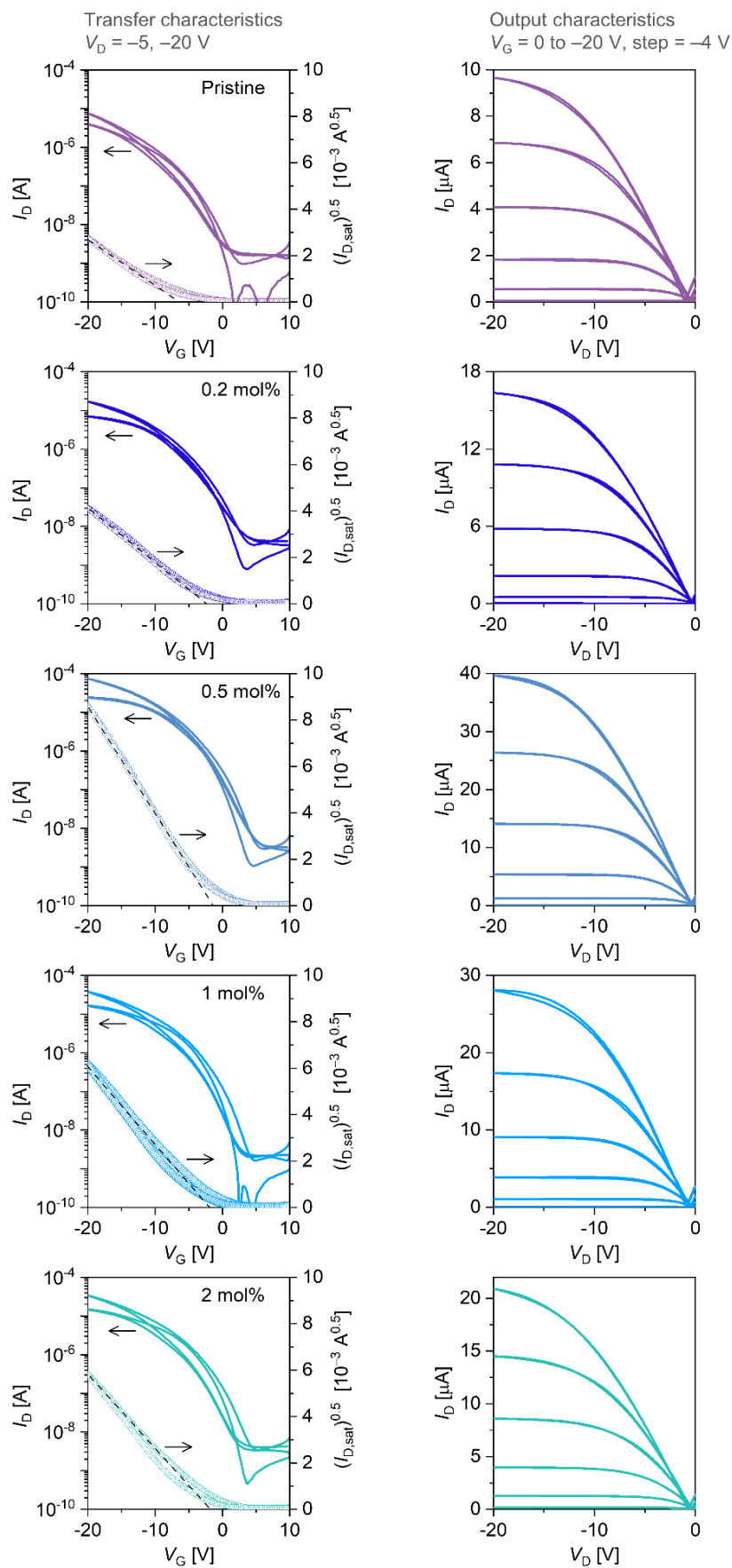

**Figure S10.** Representative transfer and output characteristics of CuSCN TFTs: 0 to 2 mol% $I_2$ doping.

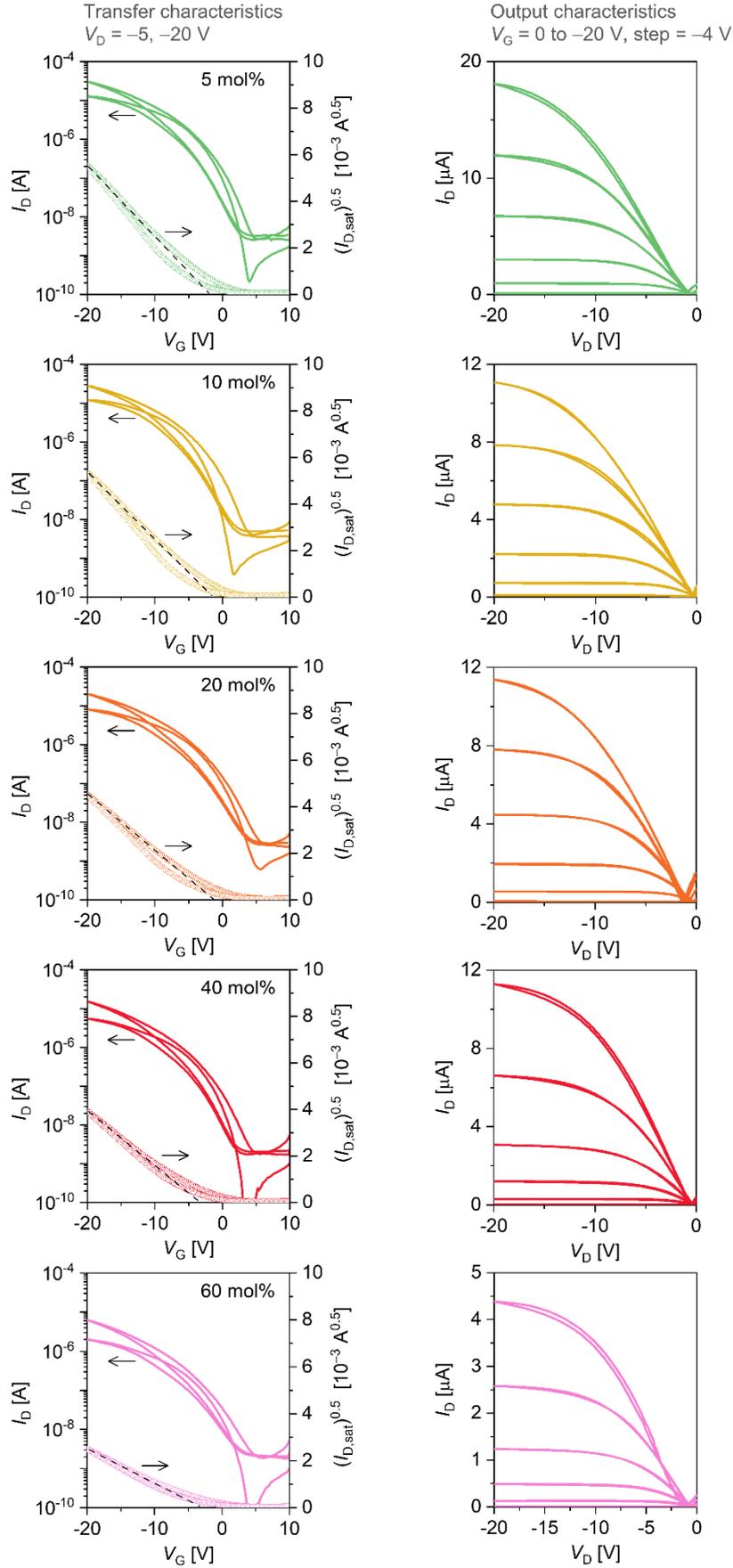

**Figure S11.** Representative transfer and output characteristics of CuSCN TFTs: 5 to 60 mol% I$_2$ doping.

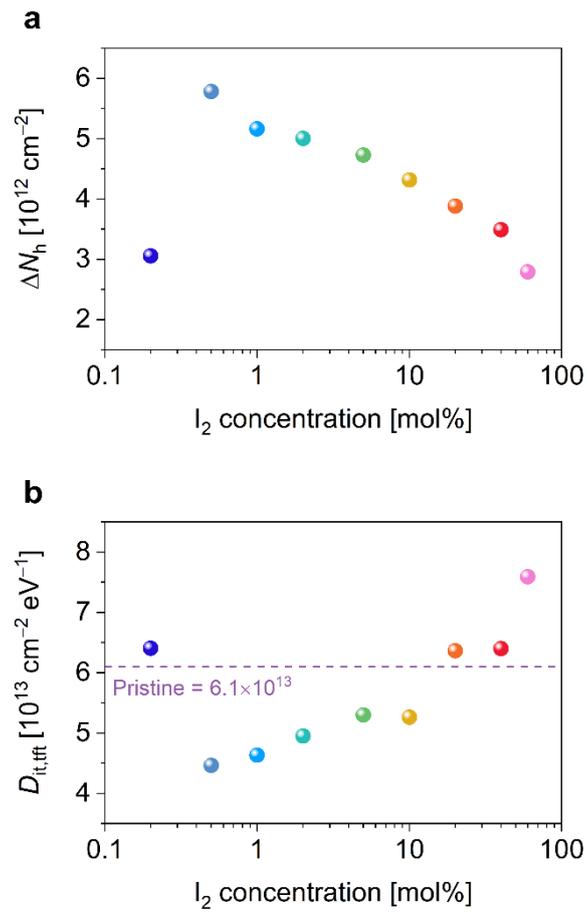

**Figure S12.** (a) The shift in the hole concentration calculated from the shift in the threshold voltage ($\Delta N_h$) and (b) interface trap density calculated from the subthreshold swing ($D_{it,tft}$) of CuSCN transistors with varying I$_2$ doping concentration.

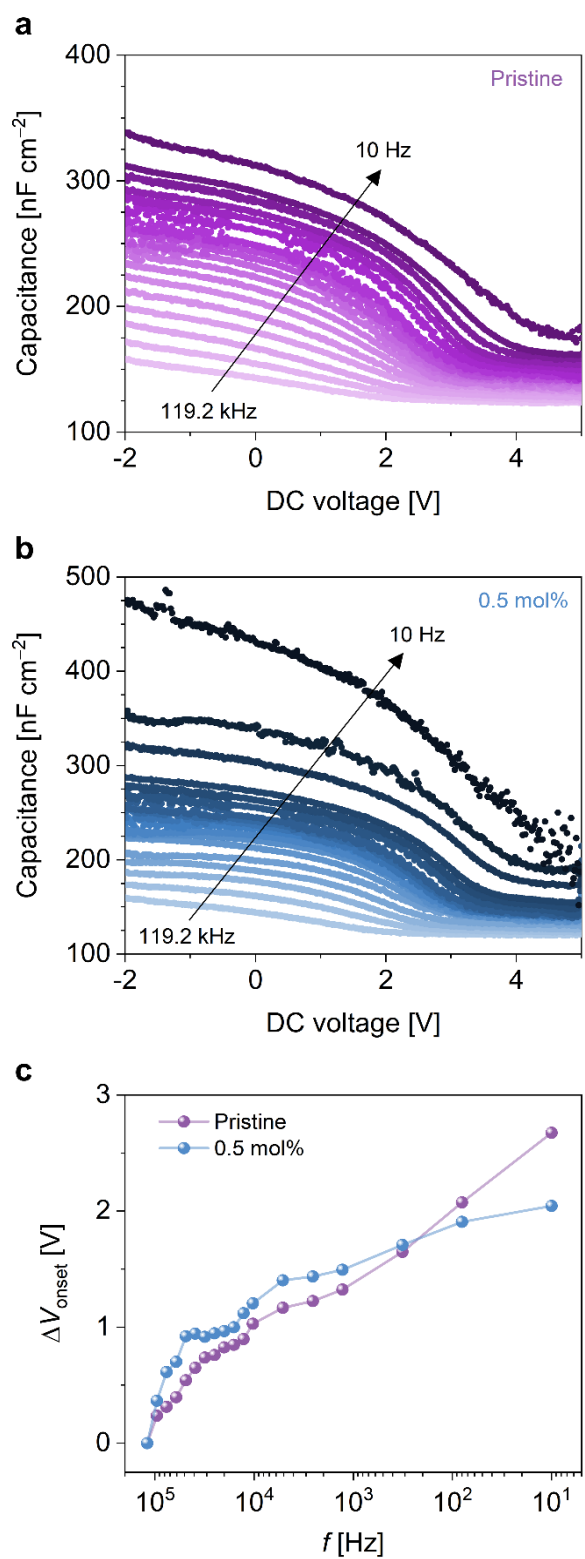

**Figure S13.** Frequency dispersion of *C-V* characteristics of metal-insulator-semiconductor (MIS) structures based on (a) pristine and (b) 0.5 mol% $I_2$-doped CuSCN. (c) The shift in the onset voltage of the *C-V* characteristics ($\Delta V_{onset}$) relative to the onset at the highest frequency (119.2 kHz).

**Table S1.** Mixing volumes of CuSCN and I₂ solutions for each doping condition. The starting concentrations of stock CuSCN and I₂ solutions in diethyl sulfide (DES) were both 10 mg ml$^{-1}$.

| Sample condition | Volume of CuSCN solution [μl] | Volume of I₂ solution [μl] | Amount of CuSCN [μmol] | Amount of I₂ [μmol] |
|---|---|---|---|---|
| Pristine | 500.0 | – | 41.11 | – |
| 0.2 mol% | 426.0 | 1.772 | 35.02 | 0.070 |
| 0.5 mol% | 424.2 | 4.430 | 34.88 | 0.175 |
| 1 mol% | 422.1 | 8.860 | 34.70 | 0.349 |
| 2 mol% | 418.3 | 17.72 | 34.39 | 0.698 |
| 5 mol% | 406.0 | 44.10 | 33.38 | 1.738 |
| 10 mol% | 385.0 | 88.20 | 31.65 | 3.475 |
| 20 mol% | 340.9 | 177.1 | 28.03 | 6.978 |
| 40 mol% | 255.5 | 354.2 | 21.01 | 13.95 |
| 60 mol% | 170.7 | 531.6 | 14.03 | 20.94 |

**Table S2:** Root-mean-square roughness ($\sigma_{rms}$) and average grain size ($D_{avg}$) from atomic force microscopy (AFM) topographic data.

| Sample | $\sigma_{rms}$ [nm] | $D_{avg}$ [nm] |
|---|---|---|
| Pristine | 4.32 | 8.38 |
| 0.5 mol% | 3.66 | 5.42 |
| 2 mol% | 4.73 | 6.14 |
| 5 mol% | 4.69 | 5.95 |
| 20 mol% | 4.35 | 5.68 |
| 40 mol% | 3.80 | 4.53 |

**Table S3.** Optical band gaps ($E_g^{opt}$) determined from Tauc plots.

| Sample | Optical band gap [eV] |
|---|---|
| Pristine | 3.97 |
| 0.5 mol% | 3.97 |
| 2 mol% | 3.97 |
| 5 mol% | 3.97 |
| 10 mol% | 3.97 |
| 20 mol% | 3.96 |
| 40 mol% | 3.96 |
| 60 mol% | 3.95 |

**Table S4.** Results from photoelectron spectroscopy (PES) using synchrotron radiation with a photon energy of 60 eV. WF: work function; $E_F$: Fermi level; $E_{vac}$: vacuum level; $E_v$: valence band edge.

| Sample | Cutoff [eV] | WF [eV] | $E_v$ below $E_F$ [eV] | $E_v$ below $E_{vac}$ [eV] |
|---|---|---|---|---|
| ITO | 55.68 | 4.32 | 3.12 | 7.44 |
| Pristine | 55.02 | 4.98 | 0.74 | 5.72 |
| 0.5 mol% | 54.85 | 5.15 | 0.72 | 5.87 |
| 5 mol% | 55.08 | 4.92 | 0.72 | 5.64 |
| 40 mol% | 55.13 | 4.87 | 0.69 | 5.56 |

**Table S5.** Device parameters of TG-BC TFTs with pristine or $I_2$-doped CuSCN films as the *p*-type semiconducting channel. The errors in the average values are give by the standard deviations. $\mu_{lin}$: linear mobility; $\mu_{sat}$: saturation mobility; $V_{th}$: threshold voltage; $S_{th}$: subthreshold swing.

| Sample | No. of devices tested | Avg $\mu_{lin}$ [$10^{-2}$ cm$^2$ V$^{-1}$ s$^{-1}$] | Avg $\mu_{sat}$ [$10^{-2}$ cm$^2$ V$^{-1}$ s$^{-1}$] | Max $\mu_{sat}$ [$10^{-2}$ cm$^2$ V$^{-1}$ s$^{-1}$] | Avg $V_{th}$ [V] | Min $|V_{th}|$ [V] | Avg $S_{th}$ [V dec$^{-1}$] | Min $S_{th}$ [V dec$^{-1}$] |
|---|---|---|---|---|---|---|---|---|
| Pristine | 10 | 1.12 ±0.07 | 1.33 ±0.07 | 1.46 | −4.64 ±0.11 | −4.49 | 2.72 ±0.23 | 2.32 |
| 0.2 mol% | 16 | 1.67 ±0.15 | 2.03 ±0.33 | 2.62 | −2.33 ±0.20 | −1.93 | 2.93 ±0.48 | 2.08 |
| 0.5 mol% | 28 | 5.65 ±0.48 | 7.12 ±0.48 | 8.06 | −0.38 ±0.13 | −0.14 | 1.99 ±0.26 | 1.31 |
| 1 mol% | 16 | 4.09 ±0.36 | 4.39 ±0.25 | 4.95 | −0.73 ±0.24 | −0.39 | 2.06 ±0.11 | 1.84 |
| 2 mol% | 23 | 3.44 ±0.23 | 3.56 ±0.18 | 3.85 | −0.85 ±0.30 | −0.48 | 2.21 ±0.51 | 1.80 |
| 5 mol% | 10 | 3.20 ±0.14 | 3.46 ±0.11 | 3.65 | −1.07 ±0.15 | −0.78 | 2.36 ±0.50 | 1.99 |
| 10 mol% | 11 | 2.96 ±0.38 | 3.32 ±0.19 | 3.72 | −1.38 ±0.23 | −1.02 | 2.34 ±0.11 | 2.19 |
| 20 mol% | 17 | 2.26 ±0.57 | 2.31 ±0.26 | 3.30 | −1.70 ±0.49 | −1.33 | 2.84 ±0.73 | 2.45 |
| 40 mol% | 18 | 1.64 ±0.20 | 1.94 ±0.08 | 2.06 | −2.00 ±0.46 | −1.48 | 2.95 ±0.28 | 2.64 |
| 60 mol% | 15 | 0.47 ±0.06 | 0.72 ±0.11 | 1.03 | −2.53 ±0.37 | −1.85 | 3.38 ±0.43 | 3.00 |

**Table S6.** Onset voltages ($V_{onset}$) of the capacitance-voltage (C-V) characteristics of the metal-insulator-semicondutor (MIS) devices at each AC frequency.

| Frequency [Hz] | Pristine $V_{onset}$ [V] | 0.5 mol% $I_2$ $V_{onset}$ [V] |
|---|---|---|
| 119,209 | 2.090 | 2.158 |
| 95,367 | 2.326 | 2.524 |
| 76,294 | 2.403 | 2.772 |
| 61,035 | 2.486 | 2.861 |
| 48,828 | 2.632 | 3.080 |
| 39,063 | 2.741 | 3.101 |
| 31,250 | 2.829 | 3.075 |
| 25,000 | 2.850 | 3.106 |
| 20,000 | 2.915 | 3.123 |
| 16,000 | 2.938 | 3.156 |
| 12,800 | 2.988 | 3.279 |
| 10,240 | 3.121 | 3.364 |
| 5,120 | 3.256 | 3.562 |
| 2,560 | 3.315 | 3.594 |
| 1,280 | 3.413 | 3.652 |
| 320 | 3.738 | 3.866 |
| 80 | 4.164 | 4.066 |
| 10 | 4.766 | 4.202 |

**Table S7.** Changes in the device parameter of CuSCN TFTs following a 3000-s negative bias stress (NBS) period at $V_G = -15$ V.

| Sample | $I_{D,on}$ ratio[a] | $I_{D,off}$ ratio[a] | $\Delta V_{on}$ [V] | $\Delta V_{th}$ [V] | $\Delta S_{th}$ [V dec$^{-1}$] |
|---|---|---|---|---|---|
| Pristine | 1.94× | 9.12× | +8.08 | +3.51 | +1.59 |
| 0.5 mol% | 1.59× | 1.95× | +2.57 | +0.70 | +0.22 |

[a] The ratio of the values after the bias-stress test compared to the initial values before the test (expressed in multiples of the initial values).